\documentclass[11pt]{JHEP3}
\pdfoutput=1
\usepackage{amsmath,amssymb,epsfig}
\usepackage{amsfonts}
\usepackage{verbatim}
\usepackage{latexsym}
\usepackage{amsthm}
\usepackage{amsbsy}
\usepackage{multirow}

\def\one{{\,\hbox{1\kern-.8mm l}}}
\newcommand{\Dslash}{\not{\hbox{\kern-4pt $D$}}}
\newcommand{\pdslash}{\not{\hbox{\kern-2pt $\partial$}}}
\newcommand{\cL}{\mathcal{L}} \newcommand{\cA}{\mathcal{A}}
\newcommand{\cN}{\mathcal{N}} \newcommand{\cO}{\mathcal{O}}

\newcommand{\Tr}{\mathrm{Tr}}

\newcommand{\SO}{\mathrm{SO}} 
\newcommand{\SU}{\mathrm{SU}} \newcommand{\U}{\mathrm{U}}
 \newcommand{\STr}{\mathrm{STr}}

\newcommand{\eg}{\emph{e.g.}\;} \newcommand{\pd}{\partial}

\newcommand{\Comment}[1]{{}}

\def\IZ{{\mathbb Z}}

\def\IR{{\mathbb R}}

\newcommand{\boldx}{\hbox{\boldmath $x$}}
\newcommand{\boldX}{\hbox{\boldmath $X$}}
\newcommand{\boldA}{\hbox{\boldmath $A$}}
\newcommand{\boldF}{\hbox{\boldmath $F$}}
\newcommand{\boldpsi}{\hbox{\boldmath $\psi$}}
\newcommand{\boldPsi}{\hbox{\boldmath $\Psi$}}
\newcommand{\boldsigma}{\hbox{\boldmath $\sigma$}}
\newcommand{\alp}{{2\pi\alpha'}}
\newcommand{\gYM}{{g_{\!\it YM}}}
\newcommand{\gYMs}{{g^2_{\!\it YM}}}

\newcommand{\bc}{\begin{center}}
\newcommand{\ec}{\end{center}}
\newcommand{\ba}{\begin{array}}
\newcommand{\ea}{\end{array}}
\newcommand{\beq}{\begin{equation}}
\newcommand{\eeq}{\end{equation}}
\newcommand{\bea}{\begin{eqnarray}}
\newcommand{\eea}{\end{eqnarray}}
\newcommand{\bmx}{\begin{pmatrix}}
\newcommand{\emx}{\end{pmatrix}}
\newcommand{\nn}{\nonumber}

\newcommand{\be}{\begin{equation}}
\newcommand{\ee}{\end{equation}}

\newcommand{\ep}{\epsilon}
\newcommand{\bep}{{\overline\epsilon}}
\newcommand{\vep}{\varepsilon}

\newcommand{\del}{\partial}
\newcommand{\half}{\frac{1}{2}}

\newcommand{\tD}{{\tilde D}}

\newcommand{\eref}[1]{Eq.\,(\ref{#1})}

\newcommand{\hA}{{\hat{A}}}

\newcommand{\hX}{{\hat{X}}}
\newcommand{\hlambda}{{\hat{\lambda}}}
\newcommand{\hD}{{\hat{D}}}
\newcommand{\hB}{{\hat{B}}}

\newcommand{\boldf}{{\bf f}}

\newcommand{\tx}{{\tilde x}}

\def\IB{\relax{\rm I\kern-.18em B}}
\def\IC{{\relax\hbox{\kern.3em{\cmss I}$\kern-.4em{\rm C}$}}}
\def\ID{\relax{\rm I\kern-.18em D}}
\def\IE{\relax{\rm I\kern-.18em E}}
\def\IF{\relax{\rm I\kern-.18em F}}
\def\II{\relax{\rm I\kern-.18em I}}
\def\IZ{\relax{\sf Z\kern-.35em Z}}
\def\Id{\relax{1\kern-.32em 1}}
\def\IG{\relax\hbox{$\inbar\kern-.3em{\rm G}$}}
\def\IR{\relax{\rm I\kern-.18em R}}
\newcommand\sfrac[2]{{\textstyle\frac{#1}{#2}}}

\newcommand\shalf{{\textstyle\frac12}}

\normalsize

\title{The Power of the Higgs Mechanism: Higher-Derivative BLG Theories}

\author{Bobby Ezhuthachan$^1$\footnote{E-mail: bobby@mri.ernet.in}\;, Sunil
  Mukhi$^2$\footnote{E-mail: mukhi@tifr.res.in}\, and Constantinos
  Papageorgakis$^2$\footnote{E-mail: costis@theory.tifr.res.in}\\~\\\it
  $^1$ Harish-Chandra Research Institute,\\
  ~ Chhatnag Rd, Jhunsi, Allahabad 211019, India \\~\\
  $^2$ Tata Institute of Fundamental Research,\\ ~ Homi Bhabha Rd,
  Mumbai 400 005, India}

\abstract{We use the novel Higgs mechanism of arXiv:0803.3218 to 
determine the leading higher-derivative corrections to the Euclidean
$\cN = 8$ Bagger-Lambert-Gustavsson field theory. The result matches
that previously found for Lorentzian 3-algebras, pointing to a
universal answer for all maximally supersymmetric 3-algebra
theories. We also comment on the extension to the
lower-supersymmetric case of ABJM theory.}

\preprint{HRI/ST/0907 \\TIFR/TH/09-06}


\begin{document}

\section{Introduction}

Recent theoretical research has yielded several classes of $(2+1)$
dimensional field theories that appear to have some relevance to
multiple M2-branes. The first breakthrough occurred when Bagger and
Lambert proposed a Lagrangian depending on an arbitrary 3-algebra and
possessing ${\cal N}=8$ supersymmetry and conformal invariance
\cite{Bagger:2006sk,Bagger:2007jr,Bagger:2007vi}. Closure of the
supersymmetry algebra was also demonstrated independently by
Gustavsson \cite{Gustavsson:2007vu}, and we will refer to these field
theories collectively as BLG theories. The ${\cal A}_4$-theory of
Ref.\,\cite{Bagger:2007vi} is a special case, characterised by an
integer Chern-Simons level $k$, and was the first to be discussed in
detail. It was conjectured in
Refs.\,\cite{Lambert:2008et,Distler:2008mk} to describe a pair of
multiple membranes at a kind of generalised orbifold. The precise
definition of this orbifold is not yet known. Moreover these theories
have no generalisation to $N>2$ membranes.

Another special case of BLG theories are the Lorentzian 3-algebra
theories, discussed in
Refs.\,\cite{Gomis:2008uv,Benvenuti:2008bt,Ho:2008ei}.\footnote{Here we
are interested primarily in the variant of these theories where a
constraint manifestly eliminates negative-norm states, as explained in
Refs.\,\cite{Bandres:2008kj,Gomis:2008be}.} Being BLG theories, these
too are Chern-Simons-like and have maximal ${\cal N}=8$ supersymmetry,
but unlike the ${\cal A}_4$-theory, they can be generalised to
arbitrary Lie algebras and they also do not have any coupling
parameter analogous to the level $k$ in the ${\cal A}_4$ case. It was
subsequently shown in Ref.\,\cite{Ezhuthachan:2008ch} that they can be
derived from maximally supersymmetric Yang-Mills theory using a
non-Abelian duality. While this makes it quite compelling that they
are correct, it is not yet clear that they provide a concretely useful
description of membranes.\footnote{However, see also
Refs.\,\cite{Cecotti:2008qs, Verlinde:2008di} for alternative
viewpoints on Lorentzian theories.}

A different class of theories are the ABJM theories
\cite{Aharony:2008ug} which are Chern-Simons-matter theories having
manifest ${\cal N}=6$ supersymmetry and an integer Chern-Simons level
$k$. These describe $N$ multiple membranes at a conventional geometric
$\mathbb C^4/\mathbb Z_k$ orbifold. It has been shown that these theories are also
described by 3-algebras, albeit with complex and non-anti-symmetric
structure constants \cite{Bagger:2008se,Schnabl:2008wj}.

All the above proposals were made to describe M-theory membranes to
lowest order in $\ell_p$, the M-theory Planck length. In each case
these worldvolume theories are related, via the novel Higgs mechanism
\cite{Mukhi:2008ux}, to the Yang-Mills theory on D2-branes, which is
the limit of the D-brane worldvolume theory where $\alpha'$
corrections are ignored. The Yang-Mills coupling depends on the vev
$v$ of a scalar in the original Chern-Simons-type theory.  It is an
interesting problem then to ask how to generalise the above proposals
to incorporate the first nontrivial corrections (which arise at order
$\alpha'^2$) to the D2-brane effective theory. As we will explain below,
these appear as  $\ell_p^3$ corrections to the corresponding
multiple membrane theories.

The initial goal of the present work is to obtain the leading higher
derivative corrections to the $\mathcal A_4$-theory. We will do
this by using the Higgs mechanism of \cite{Mukhi:2008ux} and comparing
the results with the $\alpha'^2$ corrections to the D2-brane effective
worldvolume action, which to the given order amounts to a
symmetrised-trace non-Abelian DBI theory.

To check the order of the correction we are after, note that the
Abelian DBI action for D2-branes is:
\be
\cL = -\frac{1}{(\alp)^2 \gYMs}\sqrt{-\det(g_{\mu\nu}+
\alp F_{\mu\nu})}\;,
\ee
where $F_{\mu\nu}=\del_\mu A_\nu - \del_\nu A_\mu$.
The factor of $(\gYM)^{-2}$ in front of the entire action reflects the
fact that it is a tree-level action in open string theory. 
Abelian duality is implemented by replacing the above action by the
equivalent action:
\be
\cL = \half \varepsilon^{\mu\nu\lambda} B_\mu F_{\nu\lambda}
- \frac{1}{(\alp)^2\gYMs}
\sqrt{-\det(g_{\mu\nu} + (\alp)^2 g_{YM}^4 B_\mu B_\nu)}\;.
\ee
This can be seen by integrating out $B_\mu$ whereupon one recovers the
original action.

If instead we integrate out the gauge field $A_\mu$, its equation of
motion tells us that $\del_\mu B_\nu - \del_\nu B_\mu=0$ and therefore
$B_\mu$ is the gradient of a scalar, which we write as:
\be
B_\mu \to -\frac{1}{\gYM} \del_\mu X^8\;,
\ee
where the coefficient is chosen so that the eventual kinetic term for
$X^8$ is correctly normalised.

Noting also that in static gauge
$g_{\mu\nu}=\eta_{\mu\nu} +
(\alp)^2\gYMs \del_\mu X^i\del_\nu X^i$,\footnote{The coefficient of the second term is again
determined by requiring a standard kinetic term.} and that
$(\alpha')^2\gYMs=(\alpha')^{\frac32} g_s= \ell_p^3$, we end up with
the action:
\be
\cL = -\frac{1}{(2\pi)^2\ell_p^3}\sqrt{-\det(\eta_{\mu\nu}+ 
(2\pi)^2\ell_p^3\, \del_\mu X^I\del_\nu X^I)}\sim -\half\del_\mu X^I \del^\mu
X^I + (2\pi)^2\ell_p^3\, {\cal O}(\del X)^4 + \cdots
\ee
Apparently this action depends solely on $\ell_p$. However
quantisation of flux in the original gauge theory imposes the
periodicity condition:
\be
X^8 \sim X^8 + 2\pi\gYM\;.
\ee
Therefore only in the limit $\gYM\to\infty$ (which is the same as the
M-theory limit $g_s\to\infty$) does the dependence on $\gYM$
disappear. In this limit we find an action that depends solely on
$\ell_p$ and has $\SO(8)$ invariance. This can then be interpreted as
the action for a single M2-brane.  We see that the first nontrivial
correction to this action is of order $\ell_p^3$ and this comes
multiplied by the dimension-six operator $(\del X)^4$. This fixes the
order of the corrections in which we will be interested in the
non-Abelian case as well.

The way in which we will proceed is by first presenting an ansatz for
the $\mathcal A_4$-theory action at four derivative order, in the
bi-fundamental notation of Ref.\,\cite{VanRaamsdonk:2008ft}. This is
motivated by the above considerations and the expectation that the
${\cal O}(\ell_p^3)$ corrections can be expressed in terms of
3-algebra quantities, notably the totally anti-symmetric
triple-product, with arbitrary coefficients for each possible term. We
will then match these term-by-term with the equivalent structures
arising in the D-brane effective action at order $\alpha'^2$,
following the Higgsing procedure of \cite{Mukhi:2008ux}. To this
order, the latter is given entirely by applying Tseytlin's symmetrised
trace prescription to the non-Abelian Dirac-Born-Infeld (DBI) action
\cite{Tseytlin:1997csa}, including the fermionic terms, as shown \eg
in Refs.\,\cite{Bergshoeff:2001dc, Cederwall:2001td}.

As with several discussions of the leading-order BLG $\cA_4$  and
ABJM theories, the classical action is most meaningful for large $k$
where the theory is weakly coupled and loop corrections can be
ignored. Nevertheless, it is usually written down and studied as a
function of $k$ and one hopes it has some significance even for small
$k$. In this spirit, we investigate higher-derivative corrections
keeping in mind that they too are applicable primarily in the large
$k$ regime, but the action we will obtain can  then  be
extrapolated to small $k$ with due caution.

After obtaining the result for the $\cA_4$-theory in the
bi-fundamental formulation, we move on to express the answer entirely
in terms of 3-algebra notation. We then revisit the results of
Refs.\,\cite{Alishahiha:2008rs,Iengo:2008cq}, where higher-derivative
corrections to Lorentzian 3-algebra theories were obtained by showing
that the non-Abelian duality, used in Ref.\,\cite{Ezhuthachan:2008ch},
extends to $\alpha'^2$ corrections.\footnote{In
Ref.\,\cite{Iengo:2008cq} there is a formal extension to all orders in
$\alpha'$ but the starting point used there, of a DBI-type non-Abelian
D2-brane action, is strictly correct only up to order $\alpha'^2$. For other applications of the procedure of Ref.\,\cite{Ezhuthachan:2008ch} see \cite{Kluson:2008nw}.} We
finally show that both the Euclidean and Lorentzian four-derivative
3-algebra theories are obtainable from a general four-derivative BLG
3-algebra action, upon making a Euclidean or Lorentzian choice for the
3-algebra metric.

Though we work only to order $\ell_p^3$, we conjecture that all
subsequent corrections to BLG theories can be organised in terms of
the 3-algebra triple-product. We refer to the action including all
such corrections as the ``3BI action''. Thus:
\be
S_{\textrm{3BI}} = S_{\ell_p^{\,0}} + S_{\ell_{p}^3} + \ldots \;.
\ee
We close by discussing possible generalisations of our result to
$\cN=6$ 3-algebra constructions that include the ABJM theory.

\section{Review of the novel Higgs mechanism}\label{review}
We begin with a careful review of the Higgsing procedure for the $\mathcal
A_4$-theory in the $\SU(2)\times \SU(2)$ formalism of
\cite{VanRaamsdonk:2008ft}, including the fermions. 
This will be useful to set up notation and normalisations before we
proceed to the more complicated four-derivative order.

\subsection{Higgs mechanism for the $\mathcal A_4$-theory}
The $\cA_4$-theory action is given by the expression:\footnote{This action is related
to the one in Ref.\,\cite{VanRaamsdonk:2008ft} by a
re-scaling $X\to \sqrt{\frac{k}{2\pi}}\,X$, 
$\Psi\to \sqrt{\frac{k}{2\pi}}\,\Psi$, and a
redefinition $A_\mu\to -i A_\mu$ so that the matrix-valued gauge
fields are anti-Hermitian. Our spinor and $\Gamma$-matrix notation and
conventions can be found in Appendix\,\ref{gammas}.}
\bea
\label{markform}
S_{\mathcal A_4} &=& \frac{k}{2\pi}\int d^3x\, \Tr\bigg[  -(\tilde D^\mu
  X^I)^\dagger \tilde D_\mu X^I +
i\, \bar{\Psi}^\dagger \Gamma^
{\mu}\tilde D_\mu \Psi -
\sfrac{8}{3}\, X^{IJK\dagger}\,X^{IJK} \nn\\[2mm]
&& - i\,
\bar{\Psi}^\dagger \Gamma_{IJ} [X^I, X^{J \dagger}, \Psi] + i\,
\bar \Psi \Gamma_{IJ} [X^{I\dagger}, X^J, \Psi^\dagger] 
\nn\\[2.5mm]
&& + \sfrac{1}{2}\, \varepsilon^{\mu \nu \lambda}
\Big(A^{(1)}_\mu \partial_\nu A^{(1)}_\lambda +
\sfrac{2}{3} A^{(1)}_\mu A^{(1)}_\nu A^{(1)}_\lambda
- A^{(2)}_\mu \partial_\nu A^{(2)}_\lambda
- \sfrac{2}{3} A^{(2)}_\mu A^{(2)}_\nu A^{(2)}_\lambda\Big)\bigg] \;,
\eea
where the fermions are 32-component spinors satisfying
$\Gamma^{012}\Psi = -\Psi$ and we have also defined:
\bea
\label{xijkdef}
X^{IJK} &=& X^{[I}X^{J\dagger} X^{K]}\cr
 [X^I, X^{J \dagger}, \Psi] &=& \sfrac{1}{3}\Big(X^{[I}
X^{J]\dagger}\Psi -  X^{[I} \Psi^\dagger X^{J]} + \Psi X^{[I\dagger}
X^{J]}\Big)\;.
\eea
Note that the explicit anti-symmetrisation is in the indices while leaving the
position of the $\dagger$ fixed and that the anti-symmetrised products
are defined with weight one.

In the above the indices $\mu = 0,1,2$ while $I = 1,...,8$. The
$A_\mu^{(1)}$ and $A_\mu^{(2)}$ are the $\SU(2)$ gauge fields giving rise to
the two Chern-Simons terms with equal but opposite levels $k$, which
are quantised in integer units. The
$X$'s are complex scalars obeying the reality condition:
\be
X_{a{\dot b}} = \epsilon_{ab}\,\epsilon_{{\dot b}{\dot a}}\,
X^{\dagger {\dot a}b}
\ee
and transforming in the bi-fundamental representation $({\bf 2, \bar
2})$ according to:
\be
\tilde D_\mu X^I = \partial_\mu X^I + A^{(1)}_\mu X^I 
- X^I A^{(2)}_\mu
\; .
\ee

The next step is to create linear combinations of the gauge
fields:
\bea
A_\mu &=& \sfrac{1}{2} \big(A^{(1)}_\mu + A^{(2)}_\mu\big)\nn\\ 
B_\mu &=& \sfrac{1}{2} \big(A^{(1)}_\mu - A^{(2)}_\mu\big) \; .
\eea
With these definitions the form of the bosonic part of the action
becomes:
\bea
\label{bosonic}
S^b_{\cA_4} &=& \frac{k}{2\pi}\int d^3x \,\Tr \bigg[  -( D^\mu
  X^I)^\dagger  D_\mu X^I\nn
-\sfrac{8}{3}\, 
X^{IJK\dagger}\,X^{IJK}
\nn\\
&&+ \{ B_\mu , X^I\}\{B^\mu, X^{I\dagger} \}
  + D_\mu X^{I\dagger}\,\{B_\mu, X^I\} -
 \{B^\mu , X^{I\dagger}\}\, D_\mu X^I\cr
&&+ \varepsilon^{\mu \nu \lambda} \Big(B_\mu F_{\nu\lambda} + \sfrac{1}{3}
B_\mu  B_\nu B_\lambda\Big)\bigg] \; ,\nn
\\
\eea
where we have substituted:
\be
\label{tildedee}
\tilde D_\mu X^I = D_\mu X^I - \{ B_\mu, X^I\}\;,
\ee
with:
\bea
D_\mu X^I&=& \pd_\mu X^I +  [A_\mu, X^I]\cr
F_{\mu\nu} &=& \pd_\mu A_\nu - \pd_\nu A_\mu + [A_\mu, A_\nu]\;. 
\eea
Note that the new gauge field $A_\mu$ is in the diagonal subgroup of
$\SU(2)\times \SU(2)$ and has an adjoint action on the $X$'s.

In this form for the action, one can expand the scalars into trace and
traceless parts, in a suitable basis, and also give a vev $v$ to one
of them, say $X^8$:
\bea
\label{decomposition}
X^8 &=& \shalf(v+\tilde x^8)\one + \boldx^8
\nn\\[2mm] 
X^i &=&   \shalf \tilde x^i \one + \boldx^i \;.
\eea
Here:
\be
\boldx^I = i\, x^{Ia}\, \sfrac{\boldsigma^a}{2}\;,
\ee 
with $\boldsigma^a$ the usual Pauli matrices. Recall that in
the above, $i = 1,...,7$ while $I = 1,...,8$.

In what follows we will be interested in the limit of large vev
$v\to\infty$. Having performed a decomposition of the bi-fundamental
scalars into a trace and a traceless part, we substitute back into the
action to get:
\bea
\label{subsback}
S^b_{\cA_4} &=& \frac{k}{2\pi}\int d^3x \bigg\{
-\shalf \pd^\mu \tilde x^I \pd_\mu \tilde x^I
+ \Tr \Big(
D^\mu \boldx^I D_\mu \boldx^I 
+ \sfrac{v^2}{2}  [\boldx^i, \boldx^j][\boldx^i,\boldx^j]\nn\\
&& \qquad +~ 2v B^\mu D_\mu\boldx^8
+v ^2 B^\mu B_\mu 
+ \varepsilon^{\mu \nu \lambda}
B_\mu F_{\nu\lambda} \Big)\bigg\}
+ \hbox{higher order}\;.
\eea
The higher order terms that we omitted writing in the above are the
ones that will be negligible in the final action when $v\to\infty$. We will ignore
them for now and return to them later.

One can see that after giving the vev $v$, the gauge field $B_\mu$ has
acquired a mass term by the Higgs mechanism. Moreover the
corresponding Goldstone boson that is `eaten' is $\boldx^8$, as 
is evident if we group all terms depending on $\boldx^8$ and $B_\mu$ as
follows: 
\be
v^2\left(B_\mu + \sfrac{1}{v} D_\mu \boldx^8\right)^2
+ \varepsilon^{\mu\nu\lambda}\left(B_\mu+\sfrac{1}{v}D_\mu
\boldx^8\right) 
F_{\nu\lambda}
\ee
(to obtain this form, we have added a term proportional to
$\varepsilon^{\mu\nu\lambda}D_\mu\boldx^8\, F_{\nu\lambda}$ which vanishes
by partial integration and the Bianchi identity). The shift $B_\mu \to
B_\mu - \sfrac{1}{v}D_\mu \boldx^8$ now eliminates $\boldx^8$ from the
Lagrangian.

The novel feature of this Higgs mechanism is that $B_\mu$ has no
kinetic term, therefore it can be integrated out and the effect of
this is to render the other gauge field $A_\mu$ dynamical. 
To see this, note that the equation of motion for $B_\mu$ is:
\be
\label{Asolution}
B_\mu = -\frac{1}{2 v^2} \varepsilon^{\mu\nu\lambda} F_{\nu\lambda}\;.
\ee
Eliminating $B_\mu$ from the action:
\be
  S^{b}_{\cA_4} = \frac{k}{2\pi}\int d^3x \bigg\{
- \shalf\pd^\mu \tilde x^I \pd_\mu \tilde x^I +
  \Tr \Big(\sfrac{1}{2v^2}F^{\mu\nu}F_{\mu\nu} +  D^\mu \boldx^i 
D_\mu \boldx^i + \sfrac{v^2}{2} [\boldx^i, \boldx^j][\boldx^i,\boldx^j]
\Big)\bigg\}\;.
\ee
The fields in the above action are 8 singlets $\tx^I$ along with
adjoint $\SU(2)$ scalars $\boldx^i$ and an $\SU(2)$ gauge field, all
described as matrix-valued fields in the
fundamental representation:
\be
A_\mu = i A_\mu^a\, \sfrac{\boldsigma^a}{2},\qquad 
\boldx^i =  i x^{i\,a}\,\sfrac{\boldsigma^a}{2}\;.
\ee
Extracting a factor $\sfrac{1}{v^2}$ from the action, and
re-scaling $\boldx^i\to \sfrac{1}{v}\boldx^i$ and ${\tilde x}^I \to
\sfrac{1}{v}{\tilde x}^I$, we have:
\be
  S^{b}_{\cA_4} = \frac{k}{2\pi v^2}\int d^3x \bigg\{
- \shalf\pd^\mu \tilde x^I \pd_\mu \tilde x^I +
  \Tr \Big(\sfrac{1}{2}F^{\mu\nu}F_{\mu\nu} + D^\mu \boldx^i 
D_\mu \boldx^i + \sfrac{1}{2} [\boldx^i, \boldx^j][\boldx^i,\boldx^j]
\Big)\bigg\}\;.
\ee

The last step is to combine the seven singlet scalars $\tx^i$ with the
$\SU(2)$ adjoints $\boldx^i$ to make $\U(2)$ adjoints:
\be
\label{recomposition}
\boldX^i = \sfrac{i}{2}\tx^i\one + \boldx^i\;.
\ee
This only leaves the singlet scalar $\tx^8$, which can instead be
dualised into an Abelian gauge field. This is done as follows:
\be
\label{abelian}
-\int d^3 x\; \shalf \pd^\mu \tilde x^8 \pd_\mu \tilde x^8\to
\int d^3x \;\left( -\sfrac{1}{4} F^{\mu\nu}_{\U(1)}F_{\mu\nu}^{\U(1)} +
\shalf\varepsilon^{\mu\nu\lambda}\del_\mu \tx^8 
F_{\nu\lambda}^{\U(1)} \right)\;,
\ee 
where $F_{\mu\nu}^{U(1)}$ is treated as an independent field. The
equation of motion for $F_{\mu\nu}^{U(1)}$ leads us back to the LHS. Instead,
integrating out $\tx^8$ on the RHS gives us the Bianchi identity for
$F_{\mu\nu}^{U(1)}$, solving which we have:
\be
F_{\mu\nu}^{U(1)} = \del_\mu A_\nu^{U(1)}- \del_\nu A_\mu^{U(1)}\;.
\ee
Once the Bianchi identity has been imposed, the second term on the RHS
drops out and the new Abelian gauge field combines with the $\SU(2)$
part to form a $\U(2)$ gauge field:
\bea
\boldA_\mu &=& \sfrac{i}{2}A_\mu^{U(1)}\one + A_\mu\nn\\
\boldF_{\mu\nu} &=& \sfrac{i}{2}F_{\mu\nu}^{U(1)}\one + F_{\mu\nu}\;.
\eea
Putting these ingredients together, one ends up with the
familiar-looking expression:\footnote{We will denote all fields in
D2-brane actions using bold-face symbols throughout to avoid confusion
with the $\cA_4$-theory expressions.}
\be\label{boslow}
S^{b}_{YM} = \frac{k}{2\pi v^2}\int d^3x\, \Tr
\Big(\sfrac{1}{2}\boldF^{\mu\nu}\boldF_{\mu\nu} + D^\mu \boldX^i 
D_\mu \boldX^i + \sfrac{1}{2} [\boldX^i, \boldX^j][\boldX^i,\boldX^j]
\Big)\;.
\ee
The higher-order terms that we had dropped earlier do indeed decouple
in the limit $k\to \infty$, $v\to\infty$ with the ratio
$\frac{k}{v^2}$ fixed. This is because they are of higher order in
inverse powers of $v$ but their $k$-dependence is the same as for the
leading terms.

For the fermion kinetic term and the Yukawa-type interaction with two
scalars and two fermions the procedure is now straightforward. Since
the fermions transform and decompose like the scalars:
\be
\Psi =   \shalf \tilde \psi \one + \boldpsi\quad 
\textrm{with}\qquad \boldpsi = i\psi^a
\sfrac{\boldsigma^a}{2} \;,
\ee
the trace part will reduce immediately to the required kinetic term,
while the extra term present for the traceless kinetic part, including
$\{{\bf A}^\mu , \Psi \}$, will be sub-leading in $\sfrac{1}{v}$ after the
re-scaling $\Psi\to \frac{\Psi}{v}$. The interaction term also reduces
straightforwardly upon Higgsing and by combining the trace and
$\SU(2)$ parts into anti-Hermitian fields:
\be
\boldPsi =   \sfrac{i}{2} \tilde \psi \one + \boldpsi \;,
\ee
one  gets:
\be\label{fermlow}
S^{f}_{\cA_4} = \frac{k}{2\pi v^2}\int d^3 x \; \Tr \Big[- i \bar{ \boldPsi}
\Gamma^\mu D_\mu \boldPsi -i \bar{ \boldPsi} \Gamma^i\Gamma^8 [
\boldX^i,\boldPsi] \Big] + \cO(\sfrac{1}{v})\;.
\ee
The last thing we need is to do away with the $\Gamma^8$ matrix
appearing in the second term of \eref{fermlow}. This is
straightforward if we rewrite it as:
\bea
- i \bar{ \boldPsi} \Gamma^i\Gamma^8 [
\boldX^i,\boldPsi] &=&  -i \bar{ \boldPsi} \Gamma^i(1+ \Gamma^{8}) [
\boldX^i,\boldPsi]\cr
&=&- i \bar{ \boldPsi}\sfrac{1}{\sqrt 2}(1- \Gamma^8)
\Gamma^i\sfrac{1}{\sqrt 2}(1+ \Gamma^{8}) [
\boldX^i,\boldPsi]\cr
&=&- i \bar{ \boldPsi}'
\Gamma^i [
\boldX^i,\boldPsi']\;,
\eea
where in the first step we have used $\bar\boldPsi\Gamma^i\boldPsi=0$
and in the last step we have defined:
\be
\boldPsi' = \sfrac{1}{\sqrt 2}(1+ \Gamma^{8}) \boldPsi\;.
\ee
Note that the above redefinition leaves the first term of
\eref{fermlow} invariant:
\be
- i \bar{ \boldPsi}
\Gamma^\mu D_\mu \boldPsi = - i \bar{ \boldPsi'}
\Gamma^\mu D_\mu \boldPsi'
\ee
and that the chirality condition becomes:
\be
\Gamma^{012}\boldPsi = -\boldPsi\to \Gamma^8 \boldPsi' = \boldPsi'\;.
\ee
Dropping the prime on $\boldPsi'$ for notational economy, we have:
\be
S^{f}_{\cA_4} = \frac{k}{2\pi v^2}\int d^3 x \; \Tr \Big[- i \bar{ \boldPsi}
\Gamma^\mu D_\mu \boldPsi -i \bar{ \boldPsi} \Gamma^i [
\boldX^i,\boldPsi] \Big] + \cO(\sfrac{1}{v})\;,
\ee
where the above is the action obtained by the dimensional reduction
of the fermion kinetic term of 10d YM down to 3d involving the usual
set of $\SO(9,1)$ $\Gamma$-matrices in a 32 dimensional representation, 
with $\Gamma^8$ being the $\SO(9,1)$ chirality matrix. 

Therefore, by adding the bosonic and fermionic pieces together, what we
finally recover in the limit $k\to \infty$, $v\to\infty$ with the
ratio $\frac{k}{v^2}$ fixed, 
is the action of maximally supersymmetric
$\U(2)$ Yang-Mills theory, namely the (lowest-order in $\alpha'$)
worldvolume field theory on two D2-branes. The coupling constant is
$g_{YM}^2 = 2\pi v^2/k$.

If one keeps $k$ finite while taking $v\to\infty$, the theory on the
D2-branes becomes strongly coupled. Since this belongs to the moduli
space of the $\mathcal A_4$-theory and also is, by definition, the
theory on 2 M2-branes in flat space, this amounts to a proof that
$\cA_4$ describes membranes (assuming the moduli space does not receive
significant quantum corrections, which is likely to be true given the maximal
supersymmetry). The spacetime
interpretation of the $\cA_4$-theory is not completely understood,
though some of its properties are known and it has been proposed that
it corresponds to a pair of membranes on an exotic orbifold that exists
only in M-theory \cite{Lambert:2008et,Distler:2008mk}.

\subsection{Effective Higgs rules}
\label{eHr}

Let us summarise what has happened to the theory after giving a vev to
one of the original bi-fundamental scalars, $\langle X^8\rangle =
\sfrac{v}{2}\one$: The traceless part of $X^8$ has disappeared during
the Higgsing process. The trace part $\tilde x^8$ has become an
Abelian gauge field after using the Abelian duality \eref{abelian}. Of
the two non-dynamical gauge fields, one has been integrated out while
the other has become a dynamical $\SU(2)$ in the diagonal of the
original $\SU(2)\times \SU(2)$, which combines with the above $\U(1)$
into a $\U(2)$. The fermions follow directly along similar lines. One
also has higher order terms $\cO(\sfrac{1}{v})$, which decouple in the
limit where $k\to\infty$, $v\to\infty$. Finally the scalars in the
$i=1,...,7$ directions, which were originally bi-fundamentals under
$\SU(2)\times \SU(2)$, were first separated into their trace and
trace-free parts in \eref{decomposition} and later recombined
(slightly differently) into $\U(2)$ adjoint scalars in
\eref{recomposition}.

We can summarise the above discussion into a set of effective rules
that capture the net result of the Higgsing process at this order, up
to a total derivative and $\cO(\sfrac{1}{v})$ terms. For that, we start with the
action \eref{markform} and make the following substitutions:
\begin{itemize}
\item
For the CS terms in the gauge fields:
\be
\cL_{CS} \to -\sfrac{2}{v^2}\boldf^\mu \boldf_\mu\;,
\ee
where we have defined $ \boldf^\mu =
\frac{1}{2}\varepsilon^{\mu\nu\lambda}\boldF_{\nu\lambda}$ and in
`mostly-plus' notation for the metric $(-++)$, the inverse
transformation is $\boldF_{\mu\nu} =
-\varepsilon_{\mu\nu\lambda}\boldf^\lambda$.
\item
For the scalars:\footnote{Note here that when contracting two cubic
expressions $X^{IJK} X^{IJK\dagger}$ there is an extra combinatorial
factor of 3 coming from setting any of the $\{I,J,K\}=8$. Terms of the
kind $X^{ijk}$ with $i,j,k\neq8$ will be higher order in
$\sfrac{1}{v}$ after the Higgsing and will not contribute in the large
$v$ limit.}
\bea\label{rules}
&& \!\!\!\!\tilde D^\mu X^8 \to \sfrac{1}{v}\boldf^\mu\;,~\quad 
\tilde D^\mu X^i \to \sfrac{1}{v}D^\mu\boldX^i\;,~~\quad
 X^{ij8} \to -\sfrac{1}{4v} [ \boldX^i, \boldX^j]\;,\quad
X^{ijk}\to {\cal O}\Big(\sfrac{1}{v^3}\Big)\;
\cr
&& \!\!\!\!\tilde D^\mu X^{8\dagger} 
\to -\sfrac{1}{v}\boldf^\mu\;, ~
\tilde D^\mu X^{i\dagger} \to -\sfrac{1}{v}D^\mu \boldX^i\;, 
\!\!\quad
X^{ij8\dagger}\to \sfrac{1}{4v}  [\boldX^i,\boldX^j]\;,\quad
X^{ijk\dagger}\to {\cal O}\Big(\sfrac{1}{v^3}\Big)\;.\nn\\
\eea
\item
For the fermions:
\bea
\tilde D^\mu \Psi\to \sfrac{1}{v}D^\mu \boldPsi\;, &\qquad& 
 [X^i,X^{8 \dagger},\Psi]\to -\sfrac{1}{4v} 
[\boldX^i,\boldPsi]\;, \qquad  [X^i,X^{j \dagger},\Psi]\to  
{\cal O}\Big(\sfrac{1}{v^3}\Big)\cr
\tilde D^\mu \bar\Psi \to \sfrac{1}{v} D^\mu  \bar{\boldPsi}\;, &\qquad& 
[X^{i\dagger},X^{8 },\Psi^\dagger]\to \sfrac{1}{4v} 
[\boldX^i,\boldPsi] \;, \qquad  [X^{i\dagger},X^{j },
\Psi]\to  {\cal O}\Big(\sfrac{1}{v^3}\Big)\nn\\
\eea
and $ \Gamma_{i8}\to \Gamma_i$.\footnote{When contracting the
Yukawa-type interaction with $ \Gamma_{IJ}$ there is an extra
combinatorial factor of 2 because of the $I\leftrightarrow J$
symmetry. Once again terms obtained from $[X^i,X^{\dagger j},\Psi]$
with $i,j\neq 8$ will not contribute at large $v$.}
\end{itemize}

Using these rules we can readily obtain $\U(2)$, $\cN = 8$ SYM in $(2+1)$ dimensions as in \eref{boslow}. All other terms, including those involving the gauge field, are $\cO(\sfrac{1}{v})$ and vanish in the limit $v\to\infty$, $k\to\infty$ with $\sfrac{k}{v^2}\to \textrm{fixed}$, up to a total derivative. These rules will be very useful in the following section where we consider the effect of the Higgsing process on the higher derivative terms.

\section{3BI to DBI}

We are now ready to move on to our main discussion and study the form
of the lowest non-trivial  $\ell_{p}$ corrections to the
$\cA_4$-theory.  We begin by writing down the form of the higher
derivative action at this order as a certain combination of dimension
six operators in the notation that we established in the previous
section. The main assumption we will make is that these admit an
organisation in terms of the 3-algebra product. Therefore we start
with the ansatz that the leading $\ell_{p}$ corrections take the most
general form that can arise using Euclidean 3-algebra `building blocks',
but with arbitrary coefficients. We will then use the Higgs mechanism
to uniquely determine the value of these coefficients by matching to
the leading $\alpha'$ corrections in the low-energy theory of two
D2-branes. As explained in the introduction, these corrections are
$\cO (\ell_{p}^3)$ for the ${\mathcal A_4}$-theory and
$\cO(\alpha'^2)$ for the D2-brane theory.

\subsection{Bosonic Part}
We begin with the bosonic content of the theory. Our ansatz for the
$\mathcal A_4$-theory will contain all the terms built out of
3-algebra `blocks' that are gauge/Lorentz invariant, dimension six and
lead to expressions contained in the D2-brane effective action upon
Higgsing.  However some adjustments must be made for the fact that,
unlike for the D2-brane theory, our fields $X^I$ and the corresponding
triple-product $X^{IJK}$ defined in \eref{xijkdef} are {\it complex} -- at
least in the bi-fundamental formulation of
Ref.\,\cite{VanRaamsdonk:2008ft}. As a result we first need to
re-examine the definition of symmetrised trace. We propose that this
definition be extended, for bi-fundamentals, to a symmetrisation of
the objects while keeping the daggers in their original
place. Concretely:
\be
\STr(AB^\dagger C D^\dagger)=\sfrac{1}{12}\Tr\Big[
A\big(
B^\dagger C D^\dagger +
B^\dagger D C^\dagger +
C^\dagger D B^\dagger +
C^\dagger B D^\dagger +
D^\dagger B C^\dagger +
D^\dagger C B^\dagger\big) ~+ ~\textrm{h.c.}~ \Big]
\ee
Note that this reduces to the conventional definition for Hermitian
fields, for which adding the complex conjugate is not necessary.

There is one simplification in the ${\cA_4}$-theory that should be
noted at this stage. It corresponds to an identity arising from the
low rank of the gauge group, $\SU(2)\times \SU(2)$. This identity is
straightforward to verify and states that all three possible
contractions in the $(X^{IJK})^4$ terms are proportional to each
other:
\bea\label{euclidid}
\STr\,\Big[ X^{IJK}X^{IJL\dagger}X^{MNK}X^{MNL\dagger}\Big]&=& 
2 \,\STr\, \Big[X^{IJM}X^{KLM\dagger}X^{IKN}X^{JLN\dagger}\Big]\nn\\
 &=& \sfrac13 \,\STr\,\Big[
X^{IJK}X^{IJK\dagger}X^{LMN}X^{LMN\dagger}\Big]\;.
\eea
Using this, we can write down the following general
ansatz for the ${\cal O}(\ell_p^3)$ corrections to the 
$\mathcal A_4$-theory:
\bea
\label{3BIundetermined}
(\tD X)^4: && k^2\, \STr\, \Big[{\bf a}\,
\tilde D^{\mu}X^{I}\,\tilde D_{\mu}X^{J\dagger}\, 
\tilde D^\nu X^J\, \tilde D_\nu X^{I\dagger}+ 
{\bf b}\, \tilde D^{\mu}X^{I}\,\tilde D_{\mu}X^{I\dagger}\,
\tilde D^{\nu}X^{J}\,\tilde D_{\nu}X^{J\dagger}\Big]\nn\\ 
X^{IJK} (\tD X)^3 :&& k^2\,
\varepsilon^{\mu\nu\lambda}\,\STr\,\Big[{\bf c}\,
X^{IJK} \tilde D_{\mu}X^{I\dagger}
\tilde D_{\nu}X^{J}\,\tilde D_{\lambda}X^{K\dagger} \Big]
\nn\\
(X^{IJK})^2 (\tD X)^2 :&& k^2\,
\STr\,\Big[{\bf d}\,X^{IJK}\,X^{IJK\dagger}\,\tilde D_{\mu} X^{L}\,
\tilde D^{\mu}X^{L\dagger} +
{\bf e} \, X^{IJK}\,X^{IJL\dagger}\,\tilde D_{\mu} X^{K}\,
\tilde D^{\mu}X^{L\dagger}\Big]\nn\\ 
(X^{IJK})^4 :&& k^2\,
\STr\,\Big[
{\bf f}\, \, X^{IJK}\,X^{IJK\dagger}\,X^{LMN}\,X^{LMN\dagger}\Big]\;,
\eea
where ${\bf a},{\bf b},{\bf c},{\bf d},{\bf e},{\bf f}$ are
constants which we will determine. The sum of all terms above will be
denoted $\Delta\cL$.

Note the absence of pure gauge field terms in
\eref{3BIundetermined}. Higher dimension combinations of CS terms
would break invariance under large gauge transformations. Higher
powers of the field strength would explicitly break supersymmetry, which is
expected to remain maximal in the $\ell_{p}$ expansion. 

The next step would be to Higgs the theory in
\eref{3BIundetermined} and compare with the derivative-corrected
D2-brane theory. We have already written down some `effective Higgs
rules' in Section\,\ref{eHr}. However, the rules themselves could in
principle be modified once higher-derivative corrections are
included. Fortunately, as we now show, to the lowest nontrivial order
(which is the order at which we are working) these rules in fact need
no modification.

Combining Eqs.\,(\ref{subsback}) and (\ref{3BIundetermined}), the
equation of motion for the gauge field $B_\mu$ is now of the form:
\be
\label{ab}
B_\mu= - \frac{1}{ v^2} f_\mu - \frac{1}{ v} D_\mu \boldx^8 
- \frac{\ell_{p}^3}{2 v^2} \frac{\delta(\Delta\cL)}{\delta B^\mu}\;.
\ee
We now wish to substitute this back into the $B_\mu$-dependent part of the
action:
\be
2v B^\mu D_\mu\boldx^8 + v^2 B^\mu B_\mu + \ell_p^3\,\Delta\cL(B)\;.
\ee
It is easily seen that the result is:
\be
-D^\mu\boldx^8 D_\mu\boldx^8 + \frac{1}{v^2}f^\mu f_\mu
+ \ell_p^3\, \Delta\cL\Big|_{B_\mu= -\frac{1}{v^2}f_\mu}\;.
\ee
In the process, two complicated terms at order $\ell_p^3$ have
cancelled out, considerably simplifying the computation.
The last term above is what one gets by substituting \eref{Asolution} into
$\Delta\cL$. It follows that we can apply the Higgs rules
\eref{rules} as they are, directly to the four-derivative action.
 
Through the substitutions  \eref{rules} the various pieces become:
\bea\label{semifinal}
S^b_{\bf a} &=& {\bf a}\; \left(\sfrac{k}{v^2}\right)^2
\int d^3 x\;
\STr\left[D^\mu \boldX^{i} D_\mu \boldX^j D^\nu \boldX^{i} D_\nu \boldX^j+
2 D^\mu \boldX^{i} D_\nu  \boldX^i \boldf^\mu \boldf_\nu +
\boldf^\mu \boldf_\mu \boldf^\nu \boldf_\nu\right]\cr
S^b_{\bf b} &=& {\bf b} \;
\left(\sfrac{k}{v^2}\right)^2
\int d^3 x\;
\STr\left[D^\mu \boldX^{i} D_\mu \boldX^i D^\nu \boldX^{j} D_\nu \boldX^j+
2 D^\mu \boldX^{i} D_\mu  \boldX^i \boldf^\nu \boldf_\nu +
\boldf^\mu \boldf_\mu \boldf^\nu  \boldf_\nu\right]\cr
S^b_{\bf c}&=&  {\bf c} \;\left(\sfrac{k}{v^2}\right)^2
\int d^3 x\;
\STr\left[\sfrac{3}{4} \varepsilon^{\mu\nu\lambda}  D_\mu \boldX^{i}
\boldf_\nu D_\lambda \boldX^j  \boldX^{ji} \right]\cr
S^b_{\bf d} &=& {\bf d} \;
\left(\sfrac{k}{v^2}\right)^2
\int d^3 x\;
\STr\left[\sfrac{3}{16}D^\mu \boldX^{i} D_\mu \boldX^i 
\boldX^{jk}\boldX^{jk}
+\sfrac{3}{16} \boldf^\mu \boldf_\mu \boldX^{ij}\boldX^{ij}\right]\cr
S^b_{\bf e} &=& {\bf e} \;
\left(\sfrac{k}{v^2}\right)^2
\int d^3 x\;
\STr\left[\sfrac{1}{8}D^\mu \boldX^{i} \boldX^{ij}\boldX^{kj}
D_\mu \boldX^k +
\sfrac{1}{16} \boldf^\mu \boldf_\mu \boldX^{ij}\boldX^{ij}
\right]\cr
S^b_{\bf f}&=& {\bf f} \;
\left(\sfrac{k}{v^2}\right)^2
\int d^3 x\;
\STr\left[\sfrac{9}{256} \boldX^{ij}\boldX^{ji}\boldX^{kl}
\boldX^{lk} \right]
\eea 
plus terms in $\cO(\frac{1}{v})$, where we are using $\boldX^{ij}=
[\boldX^i,\boldX^j]$.
The cancellations between the $\boldx^8$'s continue to be trivially present
at this order. This is hardly surprising if the Higgs mechanism is to
work, since these Goldstone degrees of freedom need to disappear from
the action. Putting back the factor $\ell_p^3$ in the above terms and using
\be
(2 \pi)^2 \ell_{p}^3\left(\sfrac{k}{2\pi v^2}\right)^2 =
\sfrac{(2 \pi \alpha')^2}{g_{YM}^2}
\ee%
it is now straightforward to compare with the appropriate terms coming
from the D2-brane theory. 

The precise form of the low-energy effective
action for multiple parallel D-branes is still not known to all
orders. However, up to order $\alpha'^2$ it has been explicitly
obtained using open string scattering amplitude calculations (see \eg
\cite{ Bergshoeff:2001dc,Cederwall:2001td} and
references therein) and the result agrees with Tseytlin's proposal for
a DBI action with a symmetrised prescription for the trace
\cite{Tseytlin:1997csa}. Starting from D9-branes, the prescription
requires to symmetrise over the gauge field strengths. For lower
dimensional branes, T-duality requires that this carries on to scalar
covariant derivatives and scalar commutators
\cite{Taylor:1999pr,Myers:1999ps}. This proposal fails at order $\alpha'^4$
\cite{Hashimoto:1997gm} but is good enough for our purposes.

The form of the relevant action for 2 D2-branes is given at this
order by an appropriately modified, dimensionally reduced version of
the D9-brane answer provided in \cite{Bergshoeff:2001dc}:\footnote{Note
that the coefficients here are twice their value given in
\cite{Bergshoeff:2001dc},
because the normalisation of the
trace used there is $\Tr\,( T^a T^b)=\delta^{ab}$ while we
consistently use $\Tr\, (\sfrac{\sigma^a}{2}\sfrac{\sigma^b}{2})=\half \delta^{ab}$.}
\bea
\label{bosonicDBI}
S_{\alpha'^2}^{b} &= & \sfrac{(\alp)^2}{g_{YM}^2}\int d^3x\;  \STr \Big[
  \sfrac{1}{4}\boldF_{\mu\nu}\boldF^{\nu\rho}\boldF_{\rho\sigma}\boldF^{\sigma\mu}  - \sfrac{1}{16}
    \boldF^{\mu\nu}\boldF_{\mu\nu} \boldF^{\rho\sigma}\boldF_{\rho\sigma} -
    \sfrac{1}{4}D_\mu   \boldX^{i} D^\mu  \boldX^i D_\nu  \boldX^{j} D^\nu \boldX^j\cr
 &&+~  \sfrac{1}{2}D_\mu  \boldX^{i} D^\nu \boldX^i D_\nu \boldX^{j} D^\mu \boldX^j +
\sfrac{1}{4}  \boldX^{ij} \boldX^{jk} \boldX^{kl} \boldX^{li}  -
    \sfrac{1}{16} \boldX^{ij} \boldX^{ij} \boldX^{kl} \boldX^{kl}\cr
 &&-~ \boldF_{\mu\nu}\boldF^{\nu\rho} D_\rho  \boldX^{i} D^\mu \boldX^i
  - \sfrac{1}{4} \boldF_{\mu\nu}\boldF^{\mu\nu}D_\rho \boldX^{i}
D^\rho  \boldX^i  -\sfrac{1}{8}\boldF_{\mu\nu}\boldF^{\mu\nu}  \boldX^{kl} \boldX^{kl}  \cr
 && -~\sfrac{1}{4} D_\mu \boldX^{i} D^\mu \boldX^i \boldX^{kl} \boldX^{kl}
 - \boldX^{ij} \boldX^{jk}D^\mu \boldX^{k} D_\mu \boldX^i 
-\boldF_{\mu\nu}D^\nu \boldX^{i} D^\mu \boldX^{j} \boldX^{ij} \Big] \;,
\eea 
and note that for $\U(2)$ one has the additional simplification:
\be
\STr\, \Big[\boldX^{ij}\boldX^{jk}\boldX^{kl}\boldX^{li} \Big] = \sfrac{1}{2}\STr\,
\Big[\boldX^{ij}\boldX^{ij}\boldX^{kl}\boldX^{kl} \Big]\;.
\ee
 It is then straightforward to compare the coefficients for all of these terms to
finally obtain:
\be\label{bosoniccoeff}
\begin{split}
& {\bf a } = \sfrac{1}{2} \;,\quad {\bf b } =-\sfrac{1}{4}\;,\quad
{\bf c }  =-\sfrac{4}{3} \;,\\
&{\bf d } = - \sfrac{4}{3}\;,\quad  {\bf e } = 8  \;,\quad {\bf f } =\sfrac{16}{9}\;.
\end{split}
\ee

It is important to note that the fixing of coefficients by the above
comparison is nontrivial. There are 3-algebra terms of
\eref{semifinal} that after Higgsing give rise to terms in the D2 action
\eref{bosonicDBI} coming from different index contractions (that is,
ultimately, different index contractions of the D9-brane theory before
dimensional reduction). Also in some places, two terms in the
3-algebra theory lead to the same term in the D2 action. Hence, it was
not obvious at the outset that there would be any values of the coefficients
in the above expression that would lead to the D2 theory upon
Higgsing. The fact that we find a consistent and unique set of
coefficients is therefore very satisfying.

\subsection{Fermionic Part}\label{fermionicpart}
The fermions will follow the above discussion closely. The most
general form for this part of the action at four-derivative order
is:\footnote{The issue of uniqueness is significantly more
complicated, as compared to the bosonic case, as there are many more
terms that one could write down in addition to the ones presented in
\eref{fermansatz}. However, it can be shown that these other terms can be
re-expressed by combinations already present in our
ansatz. We defer the presentation of these arguments to
Appendix\,\ref{Appunique}.}
\bea\label{fermansatz}
S^f_{\ell_{p}^3} & = & \ell_{p}^3\,k^2 \int d^3 x\; \STr
 \Big[\hat {\bf a}\; \bar \Psi^\dagger \Gamma^{IJ}[X^K,
 X^{L\dagger},\Psi]\bar \Psi^\dagger \Gamma^{KL}[X^I,
 X^{J\dagger},\Psi] + \hat {\bf b}\; \bar \Psi^\dagger \Gamma^\mu \tilde D^\nu \Psi
 \Psi^\dagger \Gamma_\nu \tilde D_\mu \Psi\cr
&&+\hat {\bf c}\; \bar\Psi^\dagger
 \Gamma^\mu[X^I,X^{J\dagger},\Psi]\bar \Psi^\dagger 
\Gamma^{IJ}\tilde D_\mu \Psi +\hat {\bf d}\; \bar{\Psi}^\dagger
\Gamma_\mu \Gamma^{IJ}\tilde D_\nu\Psi 
\tilde {D}^\mu X^{I\dagger}\tilde {D}^\nu X^J\cr
&& + \hat {\bf e} \;
\bar{\Psi}^\dagger\Gamma_\mu 
\tilde  D^\nu\Psi \tilde {D}^\mu X^{I\dagger}\tilde {D}_\nu X^I+ \hat
{\bf f} \; \bar{\Psi}^\dagger \Gamma^{IJKL}\tilde  D_\nu\Psi\; 
X^{IJK\dagger} \tilde {D}^\nu X^L \cr
&&
+ \hat {\bf g}\; \bar{\Psi}^\dagger \Gamma^{IJ}\tilde  D_\nu\Psi\;
X^{IJK\dagger}\tilde{D}^\nu  X^K
+ \hat {\bf h}\; \bar{\Psi}^\dagger \Gamma^{IJ}[X^J,X^{K\dagger},\Psi]
\tilde {D}^\mu
X^{I\dagger} \tilde{D}_\mu X^K\cr&&
+ \hat {\bf i}
\;\bar{\Psi}^\dagger\Gamma^{\mu\nu}[X^I,X^{J\dagger},\Psi]
\tilde{D}_\mu 
X^{I\dagger}\tilde{D}_\nu X^J 
+ \hat {\bf j} \;\bar{\Psi}^\dagger\Gamma_{\mu\nu}\Gamma^{IJ}[X^J,X^{K\dagger},\Psi]
\tilde{D}^\mu X^{I\dagger}\tilde{D}^\nu X^K 
\cr&&
+\hat {\bf k}\;
\bar{\Psi}^\dagger\Gamma_\mu\Gamma^{IJ}[X^K,X^{L\dagger},\Psi]
\tilde{D}^\mu X^{I\dagger}X^{JKL}
+\hat {\bf l} \; \bar{\Psi}^\dagger\Gamma_\mu[X^I,X^{J\dagger},\Psi]
\tilde{D}^\mu X^{K\dagger}X^{IJK}
\cr&&
+ \hat {\bf m}
\;\bar{\Psi}^\dagger\Gamma_\mu\Gamma^{IJKL}[X^L,X^{M\dagger},\Psi] 
X^{IJK\dagger} \tilde{D}^\mu X^M  
+ \hat
{\bf n}\;\bar{\Psi}^\dagger\Gamma_\mu\Gamma^{IJ}[X^K,X^{L\dagger},\Psi]
X^{IJK\dagger}\tilde{D}^\mu X^L
\cr&&
+\hat {\bf o}\;\bar{\Psi}^\dagger\Gamma^{IJKL}[X^M,X^{N\dagger},\Psi]
X^{IJL\dagger}X^{KMN}
+\hat {\bf p}\;\bar{\Psi}^\dagger\Gamma^{IJ}[X^K,X^{L\dagger},\Psi]
X^{IJM\dagger}X^{KLM} 
\nn\\ 
&&+\textrm{ h.c. with same coefficients}\bigg{]}\;.
\eea
It is a straightforward, but rather tedious, exercise to use the Higgs
rules and compare with the fermionic terms in the D2-brane effective
action at order $\cO (\alpha'^2)$, as given for $\U(2)$ in
\cite{Bergshoeff:2001dc}:\footnote{Again, the coefficients here are
twice their value given in \cite{Bergshoeff:2001dc},
for reasons of normalisation that we have already explained.}
\bea
S_{\alpha'^2}^f=&&\sfrac{(\alp)^2}{g_{YM}^2}\int d^3 x \;
 {\STr}\bigg{(}-\sfrac{1}{8}\bar{ \boldPsi}\Gamma^\mu 
D^\nu \boldPsi\;\bar{ \boldPsi}\Gamma_\nu D_\mu \boldPsi
-\sfrac{1}{4}\bar{ \boldPsi}\Gamma^i D^\nu \boldPsi\;\bar{ \boldPsi}
\Gamma_\nu [  \boldX^i, \boldPsi]\cr
&&
-\sfrac{1}{8}\bar{ \boldPsi}\Gamma^i [  \boldX^j,
  \boldPsi]\;\bar{ \boldPsi}\Gamma^j [  \boldX^{i}, \boldPsi] +
\sfrac{i}{2}
\bar{ \boldPsi}\Gamma_\mu D^\nu \boldPsi\; 
\boldF^{\mu\rho}\boldF_{\rho\nu}- \sfrac{i}{2}
\bar{ \boldPsi}\Gamma_\mu D^\nu \boldPsi\; 
D^\mu  \boldX^{l} D_\nu   \boldX^l\cr
&& - \sfrac{i}{2} \bar{ \boldPsi}\Gamma^i D^\nu \boldPsi\;
D^\rho   \boldX^{i} \boldF_{\rho\nu}-\sfrac{i}{2}\bar{ \boldPsi} \Gamma^i
D^\nu \boldPsi\;   \boldX^{il}  D_\nu  \boldX^l
+\sfrac{i}{2}\bar{ \boldPsi} \Gamma_\mu [  \boldX^j, \boldPsi]\;
\boldF^{\mu\rho} D_\rho  \boldX^{j }\cr
&&
-\sfrac{i}{2}\bar{ \boldPsi}\Gamma^i [  \boldX^j, \boldPsi]\; 
D^\rho  \boldX^{i} D_\rho  \boldX^{j}
+\sfrac{i}{2}\bar{ \boldPsi}\Gamma_\mu
[  \boldX^j,  \boldPsi] D^\mu   \boldX^{l}
  \boldX^{lj}+\sfrac{i}{2}\bar{ \boldPsi}\Gamma^i  [  \boldX^j, \boldPsi] 
  \boldX^{il}   \boldX^{lj}\cr
&&-\sfrac{i}{4}\bar{ \boldPsi}
\Gamma_{\mu\nu\rho} D_\sigma \boldPsi\; \boldF^{\mu\nu}\boldF^{\rho\sigma}
-\sfrac{i}{4}\bar{ \boldPsi}\Gamma_{\mu\nu\rho} [ 
  \boldX^k,  \boldPsi]\;
\boldF^{\mu\nu}D^\rho   \boldX^k+\sfrac{i}{4}\bar{ \boldPsi} \Gamma_{\mu\nu l}
D_\sigma  \boldPsi\;\boldF^{\mu\nu}D^\sigma  \boldX^l\cr
&&-\sfrac{i}{4}\bar{ \boldPsi}\Gamma_{\mu\nu l} 
[  \boldX^k, \boldPsi]\; \boldF^{\mu\nu}  \boldX^{lk}-
\sfrac{i}{2}\bar{ \boldPsi}\Gamma_{\mu j\rho} 
D_\sigma \boldPsi\; D^{\mu}   \boldX^{j}
\boldF^{\rho\sigma}-\sfrac{i}{2}\bar{ \boldPsi} \Gamma_{\mu j\rho} 
[  \boldX^k,  \boldPsi]\; D^{\mu}  \boldX^{j}D^{\rho }  \boldX^k\cr
&&
+\sfrac{i}{2}\bar{ \boldPsi}\Gamma_{\mu jl} D_\sigma \boldPsi\; 
D^{\mu}  \boldX^{j}D^{\sigma}  \boldX^l -
\sfrac{i}{2}\bar{ \boldPsi} \Gamma_{\mu j l} 
[  \boldX^k, \boldPsi]\; D^{\mu}  \boldX^{j}
  \boldX^{lk}-\sfrac{i}{4}\bar{ \boldPsi}\Gamma_{ij \rho} D_\sigma \boldPsi\;
  \boldX^{ij} \boldF^{\rho\sigma}\cr
&&
-\sfrac{i}{4}\bar{ \boldPsi}\Gamma_{ij\rho} 
[  \boldX^k, \boldPsi]\;   \boldX^{ij}D^{\rho}  \boldX^{k}
+\sfrac{i}{4}\bar{ \boldPsi}\Gamma_{ijl} D_\sigma \boldPsi\; 
  \boldX^{ij}D^{\sigma}  \boldX^l-\sfrac{i}{4}\bar{ \boldPsi} \Gamma_{ijl}
[  \boldX^k, \boldPsi]\;   \boldX^{ij}  \boldX^{lk}\bigg{)}\; .\nn\\
\eea
During the Higgs reduction and comparison of coefficients  we  use that since in 2+1 dimensions 
$\Gamma_{\mu\nu\lambda} = \varepsilon_{\mu\nu\lambda}\one_{2\times 2}$ and $\boldF_{\mu\nu}
=\varepsilon_{\mu\nu\lambda}\boldf^\lambda$, then:
\be
\STr \Big[ \bar \boldPsi  \Gamma_{\mu\nu\rho} D_\sigma \boldPsi \boldF^{\mu\nu} \boldF^{\rho\sigma}
\Big] \sim \STr \Big[ \bar \boldPsi   \Gamma_{\mu\nu\rho} D_\sigma
\boldPsi\varepsilon^{\mu\nu\kappa} \varepsilon^{\rho\sigma\lambda} \boldf_\kappa \boldf_\lambda
\Big] \sim \STr \Big[ \bar \boldPsi  D_\sigma
\boldPsi \varepsilon^{\sigma\kappa\lambda} \boldf_\kappa \boldf_\lambda
\Big] = 0
\ee
because of the STr. We also set the on-shell terms $\alpha'^2 (\Gamma_\mu D^\mu
\boldPsi + \Gamma^i[\boldX^i,\boldPsi]) $ to zero, as in \cite{Alishahiha:2008rs}, since this can be achieved by appropriate field redefinitions.
The result for the fermionic coefficients is:
\be\label{fermioncoeff}
\begin{split}
& \hat {\bf a} = - \sfrac{1}{4}\;,\quad\hat  {\bf b}=-\sfrac{1}{16}
\;,\quad\hat {\bf c}
=\sfrac{1}{4}\;,\quad \hat {\bf d} = \sfrac{i}{4} \;,\quad \hat {\bf e} =-\sfrac{i}{4}
\;,\quad \hat  {\bf f} = \sfrac{i}{6}\\
& \hat {\bf g} = -\sfrac{i}{2}\;,\quad \hat {\bf h} =-i \;,\quad \hat
{\bf i}= i \;,\quad 
\hat {\bf j}=i\;,\quad \hat {\bf k}=-2i\\
& \hat {\bf l}= 2i\;,\quad \hat {\bf m}=-\sfrac{2i}{3}\;,\quad \hat
{\bf n}=2i \;,\quad \hat {\bf o}=\sfrac{4i}{3}
\;,\quad\hat {\bf p} =4i\;.
\end{split}
\ee

\section{The four-derivative corrections in 3-algebra form}

In this section we will re-cast our results in 3-algebra language. There are several important reasons to do so. One is that we will uncover some new properties of 3-algebras, arising from the fact that at order $\ell_p^3$ we encounter traces of as many as four 3-algebra generators for the first time.

Another reason is that corrections of order $\ell_p^3$ are already
known
\cite{Alishahiha:2008rs,Iengo:2008cq}  for the special case of
Lorentzian 3-algebras. By re-writing the derivative corrections of
${\cal A}_4$-theory in terms of 3-algebra quantities, we will be able
to compare them with the results of
Refs.\,\cite{Alishahiha:2008rs,Iengo:2008cq}. Indeed, it is natural to
hope that all BLG theories (including both ${\cal A}_4$ and Lorentzian
sub-classes) originate from a common 3-algebra formulation, even
though they were obtained using completely different procedures. As we
now have all the necessary data for determining what that formulation
is, we will compare the two classes of theories explicitly. After
dealing with some issues of normalisation we will find that there is
indeed complete agreement.

Yet another reason to re-express our results in 3-algebra language is to
open the possibility of extending this investigation to the ${\cal
N}=6$ 3-algebras of Refs.\,\cite{Bagger:2008se,Schnabl:2008wj} which
encode, among other things, the ABJM field theory. In the final
subsection we will make some general comments on how this might be
done.

Let us first remind the reader of the original formulation for
BLG 3-algebra theories. Following \cite{Bagger:2006sk}, the maximally
($\cN =8$) supersymmetric 3-algebra field theory in $2+1$ dimensions
involves a set of bosonic fields $X^{Ia}, A_\mu^{~ab}$, with $I=1,...,
8$, and 32-component spinors $\Psi^a$, where $a
=1,...,\textrm{dim}_{\mathcal A}$, with $\mathrm{dim}_{\mathcal A}$
the dimension of the 3-algebra. $A_\mu^{~ab}$ is anti-symmetric in $a$
and $b$. To write the action one introduces the 4-index structure
constants $f^{abc}_{\phantom{abc}d}$ associated to the totally
anti-symmetric three-bracket:
\begin{equation}
[T^a, T^b, T^c] = f^{abc}_{\phantom{abc}d}T^d
\end{equation}
and a generalisation of the trace taken over the three-algebra
indices, which provides an appropriate 3-algebra metric:
\begin{equation}\label{trace}
 h^{ab} = \Tr(T^a T^b)\;.
\end{equation}
The  structure constants  satisfy the so-called ``fundamental identity'':
\begin{equation}
\label{funid}
f^{[abc}_{~~~~g}\,f^{e]fg}_{~~~~~d}= 0
\end{equation}
and are also completely anti-symmetric under the exchange of indices:
\beq
f^{abcd} = f^{[abcd]}\;.
\eeq
The action can then be written as:
\begin{equation}\label{BLG}
\begin{split}
S_{\textrm{BLG}} &= \int d^3 x \;  \Big[  \Tr\;\Big( -\half  \tilde D_\mu X^{I}\tilde D^\mu X^{I} +
\sfrac{i}{2} {\bar \Psi} \tilde \Dslash \Psi\\
&+\sfrac{i}{4} \bar \Psi \Gamma^{IJ}[X^{I},X^{J},\Psi]-\sfrac{1}{12} [X^{I},X^{J},X^{K}] [X^{I},X^{J},X^{K}]\Big) \\
&~~~  +\shalf\,\vep^{\mu\nu\lambda}(
\tilde A_{\mu~~b}^{~a}\; \del_\nu  A_{\lambda~~a}^{~b}
+ \sfrac23 \,A_{\mu~~b}^{~a}\tilde A_{\nu~~c}^{~b}\tilde A_{\lambda~~a}^{~c})\Big]\\
\end{split}
\end{equation}

where $\tilde A_\mu^{cd} = f_{ab}^{~~cd} A_\mu^{ab}$ and:
\begin{equation}
\tilde D_\mu X^{Ia} = \del_\mu X^{Ia} + \tilde A_{\mu~~b}^{~a}X^{Ib}\;.
\end{equation}
Note that the Tr here is the abstract 3-algebra trace defined in \eref{trace}. The fields are invariant under the gauge transformations:
\bea
\delta X^{Ia} &=& -\tilde \Lambda^{a}_{~~b}X^{Ib}\\
\delta \Psi^a & =& -\tilde \Lambda^{a}_{~~b}\Psi^b\\
\delta (\tilde A_\mu^{cd}) &=& \tilde D_\mu \tilde \Lambda^{cd} 
\eea
and the supersymmetries:
\bea
\delta X^{Ia} &=& i\,\bep\,\Gamma^I\Psi^a\\
\label{susies}
\delta \Psi^a & =& D_\mu X^{Ia}\Gamma^\mu \Gamma^I \ep +\frac16
f^a_{~~bcd}X^{Ib}X^{Jc}X^{Kd} \Gamma^{IJK}\ep\\
\delta (\tilde A_\mu^{cd}) &=& i f_{ab}^{~~\,cd}
X^{Ia}\,\bep\, \Gamma_\mu\Gamma_I
\Psi^b 
\eea
where $\Gamma_{012}\ep = \ep$ and $\Gamma_{012}\Psi^a = -\Psi^a$.

\subsection{$\cA_4$ 3-algebra theory}

For a Euclidean 3-algebra metric, $h^{ab} = \delta^{ab}$, the possible
BLG theories are the $\cA_4$-theory with $a = 1,...4$, and direct
products thereof \cite{Papadopoulos:2008sk,Gauntlett:2008uf}. Already
at the lowest (quadratic) order it is easy to see how one can convert
the above 3-algebra formulation to the bi-fundamental action of
\cite{VanRaamsdonk:2008ft} after noting a subtlety in the definition
of the trace between the two cases. Whereas $\Tr(T^a T^b) =
\delta^{ab}$ in 3-algebra notation, one has for the $\SU(2)$
generators $T^i = \sfrac{\sigma^i}{2}$, that the trace is $
\Tr(\sfrac{\sigma^i}{2}\sfrac{\sigma^j}{2}) =
\half\delta^{ij}$. Taking this into account it is straightforward to
convert \eref{BLG} into \eref{markform} and vice-versa. The powers of
$\sfrac{1}{f^2}= (\sfrac{k}{2\pi})^2$ will appear once one re-scales
the fields appropriately by $(X,\Psi)\to \sfrac{1}{\sqrt f} (X,\Psi)
$.

This is useful, since we have obtained the four-derivative action in
bi-fundamental notation and we now want to express it in 3-algebra
form. In doing so one also has to deal with evaluating the symmetrised
trace of four 3-algebra generators. Symmetry restricts its form to be:
\be\label{4Ts}
\STr\Big( T^a T^b T^c T^d \Big) = m\;h^{(ab}h^{cd)}\;,
\ee
where $m$ is a yet undetermined numerical coefficient. However, the
Lorentzian 3-algebras can help us determine the latter as
follows. Lorentzian 3-algebras include a set of generators
corresponding to a compact subgroup of the theory's whole symmetry
group. One is then free to choose them as the generators of any
semi-simple Lie algebra, \eg SU(2). In turn, tracing over the latter
leads to a flat Euclidean block in the 3-algebra metric, $h^{ij} =
\delta^{ij}$. In any four-derivative Lorentzian 3-algebra action there
will be terms with components for which the generators in \eref{4Ts}
run over this subset. In that case, and once again taking into
consideration the appropriate definition of the trace, one can
explicitly evaluate the following expression for the particular case
of SU(2):
\be
\STr\Big( T^i T^j T^k T^l \Big) = 2 \; 
\STr\Big( \sfrac{\sigma^i}{2}\sfrac{\sigma^j}{2}
\sfrac{\sigma^k}{2}\sfrac{\sigma^l}{2} \Big) =  
\sfrac{1}{4}\; \delta^{(ij}\delta^{kl)}
\ee
and this fixes $m=\sfrac{1}{4}$.

Equipped with the above fact, we can finally convert our results and
we write for the bosonic part of the $\cA_4$-theory in 3-algebra form:
\bea\label{A4algebrabosonic}
S^{b}_{\ell_{p}^3} &= & (2\pi)^2\ell_{p}^3\int d^3 x \;\STr\Big[ \sfrac14\,
\Big(\tilde D^\mu X^I \tilde D_\mu X^J  
\tilde D^\nu X^J \tilde D_\nu X^I
-\shalf \tilde D^\mu X^I \tilde D_\mu X^I \tilde D^\nu X^J \tilde D_\nu X^J\Big)\nn\\
 &&-~\sfrac{1}{6}  \,\varepsilon^{\mu\nu\lambda}\,
\Big( X^{IJK}\tilde D_\mu X^I
\tilde D_\nu X^J\tilde D_\lambda X^K\Big)\nn\\
 &&+~ \sfrac{1}{4} \,\Big(X^{IJK}X^{IJL}
\tilde D^\mu X^K\tilde D_\mu X^L 
- \sfrac{1}{6} X^{IJK} X^{IJK} \tilde D^\mu X^L \tilde D_\mu X^L 
\Big)
\nn\\
 &&+~\sfrac{1}{288} \,\Big(X^{IJK}X^{IJK} X^{LMN}X^{LMN}\Big)\Big]\;,
\eea
where now:
\be
X^{IJK} = [X^I,X^J,X^K]\;.
\ee

\subsection{Lorentzian 3-algebra theory}

In Ref.\,\cite{Alishahiha:2008rs} the equivalent four derivative terms
were constructively obtained for Lorentzian 3-algebra theories and it
was conjectured there that the ${\cal A}_4$-theory should also be
expressed in the terms of the same 3-algebra structures at four
derivative order. We will soon verify this conjecture.

Let us start by
quoting the result found in Ref.\,\cite{Alishahiha:2008rs} for the
higher-derivative corrections to Lorentzian 3-algebra theories. To
avoid confusion with the Euclidean signature theories that are
the focus of this paper, we will consistently denote all Lorentzian
3-algebra variables with a hat symbol on top. Accordingly, the
field variables in Ref.\,\cite{Alishahiha:2008rs} are
eight adjoint scalars $\hX^I$ and fermions
$\hlambda$, as well as sixteen gauge-singlet scalars and fermions
$\hX^I_\pm,\hlambda_\pm$ and a pair of gauge fields
$\hA_\mu,\hB_\mu$. 

Due to constraints, the fields $\hX^I_-,\hlambda_-$ decouple and the
fields $\hX^I_+,\hlambda_+$ are fixed to be a constant and zero,
respectively. It was shown that the bosonic part of the $\ell_p^3$
correction can be written entirely in terms of the building blocks:
\bea
\hD_\mu \hX^I &=& \del_\mu \hX^I - [\hA_\mu,\hX^I] - \hB_\mu \hX^I_+  
\nn\\~ 
\hX^{IJK} &=& \hX^I_+[\hX^J,\hX^K] + \hX^J_+[\hX^K,\hX^I] 
+ \hX^K_+[\hX^I,\hX^J]\nn\;.
\eea
To simplify formulae, we have converted the results of
Ref.\,\cite{Alishahiha:2008rs} into symmetrised-trace form. Then
Eq.(3.14) of that paper\footnote{We have corrected a few of the
coefficients.} is the sum of the following four terms (we only write
the ${\cal O}(\ell_p^3)$ corrections, dropping the leading terms):
\bea\label{Lorentzianboslead}
(\hD \hX)^4: &&
~~\,\sfrac{1}{4}\, \STr\,\Big(\hD^\mu \hX^I \hD_\mu \hX^J 
\hD^\nu \hX^J \hD_\nu \hX^I
-\shalf \hD^\mu \hX^I \hD_\mu \hX^I \hD^\nu \hX^J \hD_\nu \hX^J\Big)\nn\\
\hX^{IJK} (\hD \hX)^3: &&
-\sfrac{1}{6}\,\varepsilon^{\mu\nu\lambda}\,
\STr\,\Big( \hX^{IJK}\hD_\mu \hX^I
\hD_\nu \hX^J\hD_\lambda \hX^K\Big)\nn\\
(\hX^{IJK})^2 (\hD \hX)^2: &&~~\, \sfrac{1}{4}\, \STr\,\Big(\hX^{IJK}\hX^{IJL}
\hD^\mu \hX^K\hD_\mu \hX^L 
- \sfrac{1}{6} \hX^{IJK} \hX^{IJK} \hD^\mu \hX^L \hD_\mu \hX^L 
\Big)
\nn\\
(\hX^{IJK})^4: &&\,
~~\sfrac{1}{24}\,\STr\,\Big(\hX^{IJM}\hX^{KLM}\hX^{IKN}\hX^{JLN}-
\sfrac{1}{12} \hX^{IJK}\hX^{IJK} \hX^{LMN}\hX^{LMN}\Big)\nn\;.\\
\eea
Here, the trace is defined using $\Tr\,( T^a T^b)= \delta^{ab}$ where
$a,b$ are now adjoint Lie algebra indices.

Note that the above expression involves all possible terms one can
write down at this order using $\hD_\mu \hX^I$ and $\hX^{IJK}$ as
building blocks, with one apparent exception: The $(\hX^{IJK})^4$
terms could have contained one more distinct index contraction, namely
the one with $\hX^{IJK}\hX^{IJL}\hX^{MNK}\hX^{MNL}$. However, it is easy to
demonstrate the identity:
\bea
\label{xfourid}
\STr\Big(\hX^{IJK}\hX^{IJL}\hX^{MNK}\hX^{MNL}\Big) &=&  \STr \Big( \sfrac43
\hX^{IJM}\hX^{KLM}\hX^{IKN}\hX^{JLN}\nn\\
&& \qquad+ \sfrac19  \hX^{IJK}\hX^{IJK}
\hX^{LMN}\hX^{LMN}\Big)\;,
\eea
as a result of which only two of the three possible ${\cal
O}(\hX^{IJK})^4$ terms are independent.

\subsection{Final answer for the BLG theory}

We would finally like to recover the four-derivative action to BLG
theory for general 3-algebras. A reasonable guess would be to see whether \eref{A4algebrabosonic} provides the answer by simply  replacing the $\cA_4$ structure constants
and metric with their Lorentzian counterparts inside the
expressions. One then finds that all terms and coefficients in
\eref{Lorentzianboslead} can be readily obtained except for $\cO(\hat
X^{IJK})^4$. This discrepancy is easily traced back to the difference
between the identities obeyed by quartic powers of triple-products in
the two cases and is resolved by noticing that \eref{euclidid} is
actually a special case of \eref{xfourid}, due to the particularly
simple nature of the $\cA_4$ structure constants
$\epsilon^{abcd}$. Therefore within the class of BLG theories we are
considering, the following identity is the most general one to be
always satisfied:
\bea\label{newid}
\STr\Big(X^{IJK}X^{IJL}X^{MNK}X^{MNL}\Big)&=& \STr\Big(\sfrac43
X^{IJM}X^{KLM}X^{IKN}X^{JLN}\cr &&  \qquad+\sfrac19 X^{IJK}X^{IJK}
X^{LMN}X^{LMN}\Big)\;.
\eea
This raises the interesting question, which we leave for a future
investigation, of whether this identity is also obeyed by other
indefinite-signature BLG theories, notably those with multiple
time-like directions as discussed in
\cite{deMedeiros:2008bf,Ho:2009nk,deMedeiros:2009hf}.
If the answer turns out to be in the affirmative, then we would have
found a new relation for quartic products of structure constants that
holds for a generic $\cN=8$ 3-algebra.

With these observations we can at last write a common expression for
both ${\cal A}_4$ and Lorentzian BLG theories:    
\bea\label{3algebrabosonic}
S^b_{\textrm{BLG},\ell_{p}^3} &= & \ell_{p}^3\int d^3 x \;\STr\Big[ \sfrac14\,
\Big(\tilde D^\mu X^I \tilde D_\mu X^J  
\tilde D^\nu X^J \tilde D_\nu X^I
-\shalf \tilde D^\mu X^I \tilde D_\mu X^I \tilde D^\nu X^J \tilde D_\nu X^J\Big)\nn\\
 &&
-\sfrac{1}{6}  \,\varepsilon^{\mu\nu\lambda}\,
\Big( X^{IJK}\tilde D_\mu X^I
\tilde D_\nu X^J\tilde D_\lambda X^K\Big)\nn\\
 &&+ \sfrac{1}{4} \,\Big(X^{IJK}X^{IJL}
\tilde D^\mu X^K\tilde D_\mu X^L 
- \sfrac{1}{6} X^{IJK} X^{IJK} \tilde D^\mu X^L \tilde D_\mu X^L 
\Big)
\nn\\
 &&+\sfrac{1}{24} \,\Big(X^{IJM}X^{KLM}X^{IKN}X^{JLN}-
\sfrac{1}{12} X^{IJK}X^{IJK} X^{LMN}X^{LMN}\Big)\Big]\;.
\eea
It is very satisfactory that one can obtain the precise coefficients of \eref{bosoniccoeff} as well as \eref{Lorentzianboslead} from this expression upon specifying the 3-algebra.

Similarly we can write down the corrections for the terms including
fermions in $\cN = 8$ 3-algebra form:
\bea\label{3algebrafermionic}
S^f_{\textrm{BLG},\ell_{p}^3} & = &(2\pi)^2\ell_{p}^3 \int d^3 x\; \STr
 \Big[-\sfrac{1}{64} \bar \Psi\Gamma^{IJ}[X^K,
 X^{L},\Psi]\bar \Psi\Gamma^{KL}[X^I,
 X^{J},\Psi] -\sfrac{1}{16} \bar \Psi\Gamma^\mu \tilde D^\nu \Psi
 \Psi\Gamma_\nu \tilde D_\mu \Psi\cr
&&+\sfrac{1}{16} \bar\Psi
 \Gamma^\mu[X^I,X^{J},\Psi]\bar \Psi
\Gamma^{IJ}\tilde D_\mu \Psi +\sfrac{i}{4} \bar{\Psi}
\Gamma_\mu \Gamma^{IJ}\tilde D_\nu\Psi 
\tilde {D}^\mu X^{I}\tilde {D}^\nu X^J\cr
&& -\sfrac{i}{4} \bar{\Psi}\Gamma_\mu 
\tilde  D^\nu\Psi \tilde {D}^\mu X^{I}\tilde {D}_\nu X^I+
\sfrac{i}{24} \bar{\Psi}\Gamma^{IJKL}\tilde  D_\nu\Psi\; 
X^{IJK} \tilde {D}^\nu X^L \cr
&&
-\sfrac{i}{8} \bar{\Psi}\Gamma^{IJ}\tilde  D_\nu\Psi\;
X^{IJK}\tilde{D}^\nu  X^K
-\sfrac{i}{4} \bar{\Psi}\Gamma^{IJ}[X^J,X^{K},\Psi]
\tilde {D}^\mu
X^{I} \tilde{D}_\mu X^K\cr&&
+\sfrac{i}{4} \bar{\Psi}\Gamma^{\mu\nu}[X^I,X^{J},\Psi]
\tilde{D}_\mu 
X^{I}\tilde{D}_\nu X^J 
+\sfrac{i}{4} \bar{\Psi}\Gamma_{\mu\nu}\Gamma^{IJ}[X^J,X^{K},\Psi]
\tilde{D}^\mu X^{I}\tilde{D}^\nu X^K 
\cr&&
-\sfrac{i}{8}\bar{\Psi}\Gamma_\mu\Gamma^{IJ}[X^K,X^{L},\Psi]
\tilde{D}^\mu X^{I}X^{JKL}
+\sfrac{i}{8} \bar{\Psi}\Gamma_\mu[X^I,X^{J},\Psi]
\tilde{D}^\mu X^{K}X^{IJK}
\cr&&
-\sfrac{i}{24}\bar{\Psi}\Gamma_\mu\Gamma^{IJKL}[X^L,X^{M},\Psi] 
X^{IJK} \tilde{D}^\mu X^M  
+\sfrac{i}{8}\bar{\Psi}\Gamma_\mu\Gamma^{IJ}[X^K,X^{L},\Psi]
X^{IJK}\tilde{D}^\mu X^L
\cr&&
+\sfrac{i}{48}\bar{\Psi}\Gamma^{IJKL}[X^M,X^{N},\Psi]
X^{IJL}X^{KMN}
+\sfrac{i}{16}\bar{\Psi}\Gamma^{IJ}[X^K,X^{L},\Psi]
X^{IJM}X^{KLM} \bigg{]}\;,\nn\\
\eea
where:
\be
[X^I,X^J,\Psi] = X^I_a X^J_b \Psi_c [T^a,T^b ,T^c]\;.
\ee
 The above reduces to both \eref{fermansatz} with the
coefficients as given in \eref{fermioncoeff} and the analogous result
valid for BLG theories with Lorentzian signature as given in \cite{Alishahiha:2008rs}.

The expressions \eref{3algebrabosonic} and \eref{3algebrafermionic} are the main results of this paper.

\subsection{Towards $\cN = 6$ 3-algebra theories at four-derivative order}

It is natural to try and see whether the above can be extended to the
case of $\cN = 6$ 3-algebra theories, which include the ABJM model
\cite{Aharony:2008ug}. Finding such an extension is of great interest
as these theories have a clear spacetime interpretation in  M-theory.

One approach would be to work directly with the ABJM theory and repeat
the analysis of Section\,\ref{review}. The straightforward application
of the Higgs mechanism to the $\U(N)\times\U(N)$ ABJM theory was shown
to yield a U(N) YM action in Refs.\,\cite{Pang:2008hw, Li:2008ya}. In
ABJM the matter fields are complex, $Z^A = ( X^A + i
X^{A+4})$, where $A=1,...4$, since the R-symmetry group is
$\SU(4)\simeq \SO(6)$. In order to Higgs the theory one then gives a
vev to the real component of, say, $Z^4$. A difference between this
case and the treatment of Section\,\ref{review} is that the gauge
fields are already in $\U(N)$, as opposed to $\SU(N)$. Hence, if
everything were to work in exactly the same way as for $\cA_4$ one
would end up with an extra U(1) degree of freedom.

However, it is easy to verify that there is an extra Goldstone mode in
the problem: It is not only the {\it traceless} part of $X^4$ (the
real component of $Z^4$) that cancels out during the calculation but
also the {\it trace} part of $X^8$ (the imaginary part of
$Z^4$). Hence the number of degrees of freedom works out
right. Moreover, there is no need to perform an Abelian duality in
this context.\footnote{As far as we know this point was not noted
in Refs.\,\cite{Pang:2008hw, Li:2008ya}.}

Nevertheless, trying to construct and apply effective Higgs rules for
this case is cumbersome and becomes even more so at four-derivative
order. This is related to the fact that the ABJM matter fields are
complex with 8 real components yet reduce to real YM fields with 7
real components, hence calling for a separate treatment of $Z^{1,2,3}$
and $Z^4$. As a result, the `direct' extension is not that
straightforward and we will not attempt to carry it out here.

Another way to proceed would be to take advantage of the 3-algebra
formulation that we have just uncovered and try to generalise the
answer to the $\cN=6$ 3-algebra theories of
Refs.\,\cite{Bagger:2008se, Schnabl:2008wj}. In the latter case the
generators are complex, as are the structure constants which are
further only partially anti-symmetric under the exchange of their indices:\footnote{A different generalisation of 3-algebra theories for which the structure constants are not totally anti-symmetric was considered in \cite{Cherkis:2008qr}.}
\be
[T^a, T^b; \bar T^{\bar c}] = f^{ab\bar c}_{~~~~d} T^d\;,
\ee
with:
\be
f^{ab\bar c\bar d} = - f^{ba\bar c\bar d} \qquad\textrm{and}\qquad
f^{ab\bar c \bar d} = f^{*\bar c \bar d ab}\;,
\ee
as well as:
\be
h^{\bar a b} = \Tr (\bar T^{\bar a} T^b)\;.
\ee
The generators satisfy a complex version of the ``fundamental identity'':
\be
f^{ef\bar g}_{~~~~b}\, f^{cb\bar a}_{~~~~d} + f^{fe\bar a}_{~~~~d}\,
f^{cb\bar g}_{~~~~d}+ f^{*\bar g \bar a f}_{~~~~~\bar b}\, f^{ ce\bar
b}_{~~~~d} + f^{*\bar a \bar g e}_{~~~~~\bar b} \,f^{ cf\bar
b}_{~~~~d}=0.\;
\ee
Since we wish to be illustrative, we only focus on the bosonic piece
of the $\cN=6$ 3-algebra action, which is:
\begin{equation}\label{n6}
\begin{split}
S_{\cN=6}^b &= \int d^3 x \; \Big[ \Tr\;\Big( - \tilde D_\mu \bar
Z_{A}\tilde D^\mu Z^{A}-\sfrac{2}{3}\, \Upsilon^{CD}_B\, \bar
\Upsilon^B_{CD}\Big) \\ &~~~ +\shalf\,\vep^{\mu\nu\lambda}(
\tilde A_{\mu~~b}^{~a}\; \del_\nu  A_{\lambda~~a}^{~b}
+ \sfrac23 \,A_{\mu~~b}^{~a}\tilde A_{\nu~~c}^{~b}\tilde
A_{\lambda~~a}^{~c})\Big]\;,\\
\end{split}
\end{equation}
with
\be
\Upsilon^{CD}_{Bd} = 
f^{ab\bar c}_{~~~~ d}\,Z^C_a Z^D_b \bar Z_{B\bar c}-
\shalf\, \delta^C_B f^{ab\bar c}_{~~~~ d}\,Z^E_a Z^D_b \bar Z_{E\bar c}
+\shalf \,\delta^D_B f^{ab\bar c}_{~~~~ d}\,Z^E_a Z^C_b \bar Z_{E\bar c}\;.
\ee
Without going into all details about this theory (the interested
reader should refer to \cite{Bagger:2008se}), we would like to
highlight some relevant points. The supervariation of the fermion in
this model can be expressed as:
\be\label{supervar}
\delta \psi_{Bd} = \Dslash Z^A_d \epsilon_{AB} + \Upsilon^{CD}_{Bd}\epsilon_{CD}\;,
\ee
hence $\Upsilon^{CD}_B$ is the natural generalisation of the $\cN=8$
triple-product appearing in \eref{susies}.

Note that at lowest order the sextic scalar potential appears without
tracing any of the $\SU(4)$ indices in a given $\Upsilon$, although in
principle one could also have had terms of the type $\Upsilon^{CD}_C
\bar \Upsilon^B_{BD} $. The reason behind this is the supersymmetry of
the theory and is made manifest through \eref{supervar}. We expect
that this structure will carry on for all 3-algebra theories even when
\eref{supervar} and \eref{susies} receive $\ell_{p}^3$ corrections; in
fact, this seems necessary if we want
\eref{3algebrabosonic}-\eref{3algebrafermionic} to be invariant under
the $\cN=8$ supersymmetry variations. This suggests that all higher
derivative corrections in $\cN=6$ ought to be expressed in terms of
$\Upsilon^{CD}_B$ building blocks, in the same spirit as per our
$\cN=8$ example. 

Let us investigate how far one can go with such an ansatz. At lowest
order, the $\cN=8 $ 3-algebra action emerges as a special case of
$\cN=6$ when the structure constants are totally anti-symmetric. It is
natural to assume that the same should also hold for higher derivative
terms. Hence, \eref{3algebrabosonic} can serve as a `boundary
condition' for the higher order $\cN=6$ action. With that condition in
mind, it is easy to see that the form of the $(\tilde D Z)^4$ and
$\Upsilon (\tilde DZ)^3$ terms of interest are uniquely
determined, including the numerical coefficients.

Things start to potentially differ for $(\Upsilon)^2 (\tilde DZ)^2$ and
$(\Upsilon)^4$ terms, where one has several index contractions
available leading to the same $\cN=8$ terms as a special case. This
would suggest at first sight that it will be impossible to determine
these coefficients uniquely through Higgsing. However, we believe that
there will be generalisations of the identity \eref{newid} to $\cN=6$,
that relate terms with different index contractions. Hence, we still
hope that the Higgs mechanism will be powerful enough to also dictate
the form of the $\cN=6$ 3-algebra theory. Progress in that direction
would probably involve first understanding the origin of \eref{newid}
directly from the BLG 3-algebra point of view, as opposed to our
approach which involved studying its particular representations.

\section{Conclusions}

In this paper we have derived an extension to the BLG 3-algebra theory
at four-derivative order, which Higgses uniquely to the
four-derivative correction of the D2-brane effective worldvolume
theory. Our result applies equally to the ${\cal A}_4$ Euclidean
theory and the Lorentzian 3-algebra theory, with the latter result
having been already obtained  in
Refs.\,\cite{Alishahiha:2008rs,Iengo:2008cq}. We find it satisfying that
both classes of BLG theories have the same four-derivative
corrections, depending only on 3-algebra quantities. 

An open question raised by this investigation is to determine whether
our result applies to {\it all} BLG theories. While the ${\cal
A}_4$-theory (and its direct sums) exhausts the Euclidean signature
ones, on the Lorentzian signature side we have only looked at the
theories with one time-like direction in 3-algebra space because of
their more immediate physical relevance. However there do exist
theories with two or more time-like directions
\cite{deMedeiros:2008bf,Ho:2009nk,deMedeiros:2009hf} that we have not
covered in our analysis. If our result can be shown to apply also to
these theories then it would be truly universal for ${\cal N}=8$ BLG
theory and it might lead to a deeper understanding of the relevant
3-algebras. 

Even though the Higgs mechanism constrains the four-derivative BLG
action uniquely, it is crucial to explicitly check that it is
invariant under the set of supersymmetry transformations. This should
be done both for the Lorentzian as well as the ${\cal A}_4$
cases. For the former there is a constructive method to
carry out the supersymmetry analysis, starting with the corresponding
analysis for D2-branes and using the methods of
Refs.\,\cite{Ezhuthachan:2008ch,Alishahiha:2008rs}. For the latter,
one has to use the Higgs mechanism as a guide. However ultimately the
results for the two cases should converge into a common formula valid
for all BLG theories or at least for the two classes of BLG theories
studied here.

One would like to extend our method to find derivative corrections
involving more than four derivatives (equivalently, to order higher
than $\ell_p^3$). On the Lorentzian side, Ref.\,\cite{Iengo:2008cq} has
proposed an action to all orders in $\ell_p$ that reduces to the
action of Refs.\,\cite{Taylor:1999pr,Myers:1999ps} after Higgsing. Something
similar can surely be done for the ${\cal A}_4$-theory. However it is
important to keep in mind that the D-brane action of
Refs.\,\cite{Taylor:1999pr,Myers:1999ps} works for certain purposes such
as finding classical solutions, but cannot be considered correct as
far as generating string amplitudes is concerned (since it is known
that the symmetrised trace prescription does not work beyond four
derivatives).

Our findings also support the idea of a spacetime realisation for the
$\cA_4$-theory. In Refs.\,\cite{Lambert:2008et,Distler:2008mk} a
proposal for such an interpretation was made in terms 2 M2-branes on a
yet unknown `orbifold' of M-theory, dubbed as an ``M-fold'', which
preserves maximal, $\cN=8$, supersymmetry and has a moduli space
$(\mathbb R^8\times\mathbb R^8)/\mathbb D_{2k}$, where $\mathbb
D_{2k}$ is the dihedral group of order $4k$. The fact that we are able
to find an ${\cal O}(\ell_p^3)$ correction to the action, from which
one can recover the precise $\alpha'^2$ corrections to the D2-brane
theory by Higgsing, is encouraging and strongly suggests that such a
spacetime description should exist.

It has to be noted in this context that one expects $\ell_{p}$
corrections in M-theory to give rise to both the $\alpha'$ as well as
$g_s$ corrections in string theory. While the action we have found
reproduces the first $\alpha'$ correction by construction, it is not
clear what part of the corresponding $g_{s}$ correction (if any) it
reproduces, since in general $g_{s}$ corrections are expected to be
non-local. It would therefore be nice to understand which aspects of
the membrane dynamics are captured by the higher derivative action
that we have constructed. As indicated in the Introduction,
at large $k$ one expects to be safe because the ${\cal A}_4$-theory
 is weakly coupled, so this caveat only applies when we take $k$
small.\footnote{We would like to acknowledge the participants of 
the Indian Strings Meeting in Pondicherry (ISM08) in December 2008
for useful comments on this point.}

Finally we discussed possible generalisations of our result to $\cN=6$
3-algebras and the ABJM theory. Here we did not find a complete
result, but have sketched how one can approach the problem. It is of
considerable interest to explicitly pursue this direction for two
reasons: these models have a well-understood spacetime interpretation
at finite $k$ in terms of membranes at a geometric orbifold, and one can also
use them to perform precision calculations at large $k$ via the
$\textrm{AdS}_4/\textrm{CFT}_3$ correspondence.  We hope to report on
this in more detail in the future.

\acknowledgments 
We are grateful to Eduard Antonyan, David Berman, Rajesh Gopakumar, Neil Lambert,
Manavendra Mahato, Shiraz Minwalla, Akitsugu Miwa, Sanjaye Ramgoolam, Ashoke Sen,
Arkady Tseytlin and Mark van Raamsdonk for helpful discussions and comments.

\begin{appendix}

\section{A note on spinor conventions}\label{gammas}

Throughout this paper we have used 32-component spinors $\Psi$ for our
3-algebra theories. These are acted upon by $\Gamma$-matrices of
$\SO(10,1)$. The latter can be arranged in terms of the following
$\SO(2,1)\times \SO(8)$ decomposition:
\be\label{decoposition}
 \Gamma^M = \{\hat \gamma^\mu 
\otimes \gamma^9 , \one_{2\times 2}\otimes \gamma^I   \}\;,
\ee
where $\mu = 0,1,2$ and $I = 1,... ,8 $, $\gamma^9 = \gamma^1\ldots\gamma^8$ is the $\SO(8)$ chirality
matrix,  while the $\Gamma$-matrices
satisfy the Clifford algebra $\{ \Gamma^M, \Gamma^N\} = 2
\eta^{MN}$. The $\SO(2,1)$ $\hat \gamma$-matrices obey the following
identities, defined with weight one:
\bea
\hat \gamma_{\mu\nu} &=& \sfrac{1}{2}
(\hat \gamma_\mu \hat \gamma_\nu - \hat \gamma_\nu
\hat \gamma_\mu)\cr
\hat \gamma_{\mu\nu\lambda} &=& \hat \gamma_\mu 
\hat \gamma_\nu \hat \gamma_\lambda -
\hat \gamma_\mu \eta_{\nu\lambda} + \hat \gamma_\nu \eta_{\mu\lambda}-
\hat \gamma_\lambda \eta_{\mu\nu}\cr
\hat \gamma_\mu \hat \gamma_{\nu\lambda}&=& \hat \gamma_{\mu\nu\lambda}+
\hat \gamma_\lambda\eta_{\mu\nu} - \hat \gamma_\nu\eta_{\mu\lambda}\cr
\hat \gamma_{\nu\lambda}\hat \gamma_\mu&=
& \hat \gamma_{\nu\lambda\mu} + \hat \gamma_\nu
\eta_{\mu\lambda} - \hat \gamma_\lambda\eta_{\mu\nu}\cr
\varepsilon_{\mu\nu\lambda} \one_{2\times 2}&=& \hat \gamma_{\mu\nu\lambda}\cr 
\varepsilon_{\mu\nu\lambda}\hat \gamma^\lambda &=& \hat \gamma_{\mu\nu}\cr
\varepsilon_{\rho\sigma\nu}\hat \gamma^{\nu\mu} &= &
2\delta^\mu_{[\sigma}\hat \gamma_{\rho]}\cr
\hat \gamma_0 \hat \gamma_0 &=& -1\;.
\eea 
Moreover, the 3-algebra spinors are Goldstinos of the symmetry
breaking \eref{decoposition} and hence obey the following chirality
condition
\cite{Bagger:2007vi,VanRaamsdonk:2008ft}:
\be
 \Gamma_{012}\Psi = - \Psi \;,
\ee 
which translates to: 
\bea
 \Gamma_{012}\Psi 
&=&( \hat \gamma_{012}\otimes \gamma_9) \Psi\cr
&=&( \varepsilon_{012}\one_{2 \times 2} \otimes \gamma_9) \Psi\cr
&=&-(\one_{2 \times 2}
\otimes \gamma_9) \Psi\cr
&\equiv& -  \Gamma_9 \Psi\cr
& =& -\Psi \cr
\Rightarrow  \Gamma_9 \Psi &=& \Psi\;.
\eea
We are working with conventions where $\varepsilon_{012} = -1$, that is
$\{\hat\gamma_0,\hat\gamma_1,\hat\gamma_2\} = \{\sigma_1,-i \sigma_2,
\sigma_3\}$ and $\sigma$ the usual Pauli matrices. One can then use
$\Gamma_9$ to get the 11d identities:
\bea
\Gamma_{\mu\nu} &=& \sfrac{1}{2}(\Gamma_\mu
\Gamma_\nu - \Gamma_\nu 
\Gamma_\mu)\cr
\Gamma_{\mu\nu\lambda} &=& \Gamma_\mu \Gamma_\nu \Gamma_\lambda -
\Gamma_\mu \eta_{\nu\lambda} + \Gamma_\nu \eta_{\mu\lambda}-
\Gamma_\lambda \eta_{\mu\nu}\cr
\Gamma_\mu \Gamma_{\nu\lambda}&=& \Gamma_{\mu\nu\lambda}+
\Gamma_\lambda\eta_{\mu\nu} - \Gamma_\nu\eta_{\mu\lambda}\cr
\Gamma_{\nu\lambda}\Gamma_\mu&=&
\Gamma_{\nu\lambda\mu} + \Gamma_\nu 
\eta_{\mu\lambda} - \Gamma_\lambda\eta_{\mu\nu}\cr
\hat\varepsilon_{\mu\nu\lambda} &\equiv& \varepsilon_{\mu\nu\lambda} \Gamma^9 =
\Gamma_{\mu\nu\lambda}\cr  
\hat \varepsilon_{\mu\nu\lambda}\Gamma^\lambda &=& \Gamma_{\mu\nu}\cr
\hat\varepsilon_{\rho\sigma\nu} \Gamma^{\nu\mu} &= &
2\delta^\mu_{[\sigma}\Gamma_{\rho]}\cr
\Gamma_9 \Gamma_9 &=& 1\cr
\Gamma_0 \Gamma_0 &=& -1\;.
\eea 
We have implemented the above identities in
Subsection\,\ref{fermionicpart}.  Note that while $ \Gamma^9$
anti-commutes with the $ \Gamma^i$'s, it commutes with the $
\Gamma^\mu$'s.

\section{Explicit Higgsing of the fermionic terms}

Here we give a complete list for the explicit Higgsing of the
fermionic terms that we presented in \eref{fermansatz}. Applying the
Higgs rules of Section\,\ref{eHr} these give:
\bea
 \hat {\bf a}\; \bar \Psi^\dagger \Gamma^{IJ}[X^K,
 X^{L\dagger},\Psi]\bar \Psi^\dagger \Gamma^{KL}[X^I,
 X^{J\dagger},\Psi]&\to&  \sfrac{\hat {\bf a}}{v^4}  \sfrac{1}{2} \bar
 \boldPsi \Gamma^{i} [\boldX^j,\boldPsi] \bar \boldPsi
 \Gamma^{j}[\boldX^i,\boldPsi]\cr 
 \hat {\bf b}\; \bar \Psi^\dagger \Gamma^\mu \tilde D^\nu \Psi\bar
 \Psi^\dagger \Gamma_\nu \tilde D_\mu \Psi&\to& 2\sfrac{\hat
   {\bf b}}{v^4}\bar \boldPsi \Gamma^\mu D_\nu \boldPsi \bar
 \boldPsi \Gamma^\nu D_\mu \boldPsi\cr 
\hat {\bf c}\; \bar\Psi^\dagger
 \Gamma^\mu[X^I,X^{J\dagger},\Psi]\bar \Psi^\dagger 
\Gamma^{IJ}\tilde D_\mu \Psi&\to&  - \sfrac{\hat
  {\bf c}}{v^4}\bar \boldPsi \Gamma^\mu [\boldX^i,\boldPsi] \bar \boldPsi
\Gamma^{i} D_\mu \boldPsi\cr
 \hat {\bf d}\; \bar{\Psi}^\dagger \Gamma_\mu \Gamma^{IJ}\tilde D_\nu\Psi
\tilde {D}^\mu X^{I\dagger}\tilde {D}^\nu X^J&\to&  \sfrac{\hat
  {\bf d}}{v^4} \bar \boldPsi
\Gamma^{\rho\sigma}\Gamma^{i}D_\nu \boldPsi D^\nu \boldX^i 
\boldF_{\rho\sigma}\cr&&  + 2\sfrac{\hat {\bf d}}{v^4}\bar \boldPsi
\Gamma^{\mu} \Gamma^{ i j}D_\nu 
\boldPsi D_\mu \boldX^i D^\nu \boldX^j\cr &&
+ 2 \sfrac{\hat {\bf
    d}}{v^4} \bar \boldPsi \Gamma^{\mu\rho}\Gamma^{i} 
 D^\sigma \boldPsi D_\mu \boldX^i \boldF_{\rho\sigma}
\cr 
&&- 2 \sfrac{\hat {\bf
    d}}{v^4} \bar \boldPsi \Gamma^{i} 
 D^\sigma \boldPsi D^\rho \boldX^i \boldF_{\rho\sigma}
\cr
&& + \sfrac{\hat{\bf d}}{v^4} \bar \boldPsi \Gamma^{\rho\sigma} \Gamma_\mu \Gamma^{i}
\Dslash 
\boldPsi D^\mu \boldX^i \boldF_{\rho\sigma}
\cr 
\hat {\bf e} \; \bar{\Psi}^\dagger\Gamma_\mu
\tilde  D^\nu\Psi \tilde {D}^\mu X^{I\dagger}\tilde {D}_\nu X^I&\to& -
2 \sfrac{\hat
  {\bf e}}{v^4} \bar \boldPsi \Gamma_\mu D^\nu \boldPsi 
  \boldF^{\mu\rho}\boldF_{\rho\nu}\cr 
&& + 2\sfrac{\hat {\bf e}}{v^4} \bar \boldPsi \Gamma_\mu D^\nu
\boldPsi D^\mu \boldX^i D_\nu \boldX^i 
\cr
&&- \sfrac{\hat{\bf e}}{v^4} \bar \boldPsi \Dslash \boldPsi
\boldF^{\rho\sigma}\boldF_{\rho\sigma}\cr
 \hat {\bf f} \; \bar{\Psi}^\dagger \Gamma^{IJKL}\tilde  D_\nu\Psi\;
X^{IJK\dagger} \tilde {D}^\nu X^L&\to&  \sfrac{\hat
  {\bf f}}{v^4}\sfrac{3}{2}\bar \boldPsi \Gamma^{ijl}D_\nu
\boldPsi \boldX^{ij}D^\nu \boldX^l
\cr
 \hat {\bf g}\; \bar{\Psi}^\dagger \Gamma^{IJ}\tilde  D_\nu\Psi\;
X^{IJK\dagger}\tilde{D}^\nu  X^K&\to&  \sfrac{\hat {\bf g}}{v^4}
\bar \boldPsi \Gamma^{i} D_\nu \boldPsi \boldX^{ij}D^\nu \boldX^j\cr
&&+\sfrac{\hat {\bf g}}{v^4}\sfrac{1}{2} \bar \boldPsi
\Gamma^{ij}\Gamma^{\rho}D^\sigma\boldPsi \boldX^{ij}\boldF_{\rho\sigma} 
\cr
&& -\sfrac{\hat {\bf g}}{4v^4} \bar \boldPsi
\Gamma^{ij}\Gamma^{\rho\sigma}\Dslash \boldPsi
\boldX^{ij}\boldF_{\rho\sigma}
\cr
 \hat {\bf h}\; \bar{\Psi}^\dagger \Gamma^{IJ}[X^J,X^{K\dagger},\Psi]
\tilde {D}^\mu
X^{I\dagger} \tilde{D}_\mu X^K&\to&\sfrac{1}{2}  \sfrac{\hat {\bf h}}{v^4} \bar \boldPsi
\Gamma^{i} [\boldX^k,\boldPsi]D_\mu \boldX^i D^\mu \boldX^k\cr
&&+\sfrac{\hat {\bf h}}{v^4}\sfrac{1}{4}\bar \boldPsi
\Gamma^{\mu\rho\sigma}[\boldX^i,\boldPsi]D_\mu
\boldX^i \boldF_{\rho\sigma}
\cr
&& - \sfrac{\hat {\bf h}}{v^4}\sfrac{1}{4} \bar \boldPsi
\Gamma_{\mu\rho\sigma} \Gamma^i (\Gamma^j[\boldX^j,\boldPsi])D^\mu
\boldX^i \boldF^{\rho\sigma}\cr
&& - \sfrac{\hat {\bf h}}{v^4}\sfrac{1}{4} \bar \boldPsi
 (\Gamma^{j}[\boldX^j,\boldPsi])\boldF_{\rho\sigma} \boldF^{\rho\sigma}
\cr
  \hat {\bf i}
  \;\bar{\Psi}^\dagger\Gamma^{\mu\nu}[X^I,X^{J\dagger},\Psi]
  \tilde{D}_\mu 
X^{I\dagger}\tilde{D}_\nu X^J &\to&  \hat
  {\bf i}\bar \boldPsi \Gamma_\rho[\boldX^i,\boldPsi]D_\mu \boldX^i \boldF^{\rho\mu}
\cr
 \hat {\bf j}
 \;\bar{\Psi}^\dagger\Gamma_{\mu\nu}\Gamma^{IJ}[X^J,X^{K\dagger},\Psi] 
\tilde{D}^\mu X^{I\dagger}\tilde{D}^\nu X^K 
&\to& +\sfrac{\hat {\bf j}}{v^4}\sfrac{1}{2} \bar \boldPsi
\Gamma^{\mu\nu}\Gamma^{ i} [\boldX^k,\boldPsi] D_\mu \boldX^i D_\nu \boldX^k\cr
&&+\sfrac{\hat {\bf j}}{v^4}\sfrac{1}{2} \bar \boldPsi
\Gamma^\sigma[\boldX^i, \boldPsi] 
D^\rho \boldX^i \boldF_{\rho\sigma}
\cr
&& -\sfrac{\hat {\bf j}}{v^4}\sfrac{1}{2}\bar \boldPsi  \Gamma^\rho
 \Gamma^i (\Gamma^j[\boldX^j, \boldPsi])D^\sigma \boldX^i \boldF_{\rho\sigma}
\cr
\hat {\bf k}\;
\bar{\Psi}^\dagger\Gamma_\mu\Gamma^{IJ}[X^K,X^{L\dagger},\Psi]
\tilde{D}^\mu X^{I\dagger}X^{JKL}&\to& \sfrac{\hat
  {\bf k}}{v^4}\sfrac{1}{4}\bar \boldPsi \Gamma^{\mu}\Gamma^{ ij}[\boldX^k,
\boldPsi]  D_\mu
\boldX^i \boldX^{jk}
\cr
&& -\sfrac{1}{8}\sfrac{\hat {\bf k}}{v^4} \bar \boldPsi
\Gamma^{\rho\sigma}\Gamma^{j} [\boldX^k, \boldPsi]\boldX^{jk}\boldF_{\rho\sigma}\cr
\hat {\bf l} \; \bar{\Psi}^\dagger\Gamma_\mu[X^I,X^{J\dagger},\Psi]
\tilde{D}^\mu X^{K\dagger}X^{IJK}
&\to&-\sfrac{\hat  {\bf l}}{v^4}\sfrac{1}{4} \bar \boldPsi
\Gamma_\mu[\boldX^i, \boldPsi] D^\mu
\boldX^k \boldX^{ik}\cr
\hat {\bf m}
\;\bar{\Psi}^\dagger\Gamma_\mu\Gamma^{IJKL}[X^L,X^{M\dagger},\Psi] 
X^{IJK\dagger} \tilde{D}^\mu X^M &\to& \sfrac{\hat
  {\bf m}}{v^4} \sfrac{3}{8}\bar \boldPsi \Gamma^{\rho\sigma i}
 [\boldX^j, \boldPsi] \boldX^{ij}\boldF_{\rho\sigma}\cr &&
 -\sfrac{\hat {\bf m}}{v^4}\sfrac{3}{16}\bar \boldPsi
\Gamma^{\rho\sigma}\Gamma^{ij}(\Gamma^l [\boldX^l,
\boldPsi])\boldX^{ij} \boldF_{\rho\sigma}
\cr
\hat
{\bf n}\;\bar{\Psi}^\dagger\Gamma_\mu\Gamma^{IJ}[X^K,X^{L\dagger},\Psi]
X^{IJK\dagger}\tilde{D}^\mu X^L
&\to& - \sfrac{\hat
  {\bf n}}{v^4}\sfrac{1}{8} \bar \boldPsi \Gamma^{\rho\sigma}\Gamma^{j} [\boldX^k,
\Psi]\boldX^{jk}\boldF_{\rho\sigma}\cr
&&-\sfrac{\hat {\bf n}}{v^4}\sfrac{1}{8} \bar \boldPsi
\Gamma^{\mu}\Gamma^{ ij}[\boldX^l, \boldPsi]\boldX^{ij}D_\mu \boldX^l
\cr
\hat {\bf o}\;\bar{\Psi}^\dagger\Gamma^{IJKL}[X^M,X^{N\dagger},\Psi]
X^{IJL\dagger}X^{KMN}&\to&-\sfrac{\hat
  {\bf o}}{v^4}\sfrac{3}{16}\bar \boldPsi \Gamma^{ijk}[\boldX^m,
\boldPsi] \boldX^{ij} \boldX^{km}
 \cr
\hat {\bf
  p}\;\bar{\Psi}^\dagger\Gamma^{IJ}[X^K,X^{L\dagger},\Psi]
X^{IJM\dagger}X^{KLM} 
&\to& -\sfrac{\hat
  {\bf p}}{v^4}\sfrac{1}{8}\bar \boldPsi \Gamma^{j}[\boldX^k,
\boldPsi]\boldX^{jm}\boldX^{km} \;,
\eea
where on each right hand side of the above we have included a factor of $2$
contribution from also taking into account the Higgsing of the
Hermitian conjugates. We have made heavy use of the $\Gamma$-matrix
identities from Appendix\,\ref{gammas}.

Note that terms containing parts of the on-shell terms, $\alpha'^2(
\Gamma^\mu D_\mu \boldPsi + \Gamma^{i}[X^i, \boldPsi])$, will combine
and cancel out:
\bea
&& - \sfrac{\hat{\bf e}}{v^4} \bar \boldPsi \Dslash \boldPsi
\boldF^{\rho\sigma}\boldF_{\rho\sigma} - \sfrac{\hat {\bf
    h}}{v^4}\sfrac{1}{4} \bar \boldPsi 
 (\Gamma^j[\boldX^j,\boldPsi])\boldF_{\rho\sigma} \boldF^{\rho\sigma}
 = 0\\
&&- \sfrac{\hat {\bf g}}{4v^4} \bar \boldPsi
\Gamma^{ij}\Gamma^{\rho\sigma}\Dslash \boldPsi
\boldX^{ij}\boldF_{\rho\sigma} -\sfrac{\hat {\bf m}}{v^4}\sfrac{3}{16}\bar \boldPsi
\Gamma^{\rho\sigma}\Gamma^{jk}(\Gamma^l [\boldX^l,
\boldPsi])\boldX^{jk} \boldF_{\rho\sigma} =0\\
&&  \sfrac{\hat{\bf d}}{v^4} \bar \boldPsi \Gamma^{\rho\sigma} \Gamma_\mu \Gamma^i \Dslash
\boldPsi D^\mu \boldX^i \boldF_{\rho\sigma}- \sfrac{\hat {\bf
    h}}{v^4}\sfrac{1}{4} \bar \boldPsi 
\Gamma_{\mu\rho\sigma} \Gamma^i (\Gamma^j[\boldX^j,\boldPsi])D^\mu
\boldX^i \boldF^{\rho\sigma}\cr && \qquad \qquad -\sfrac{\hat {\bf
    j}}{v^4}\sfrac{1}{2}\bar 
\boldPsi \Gamma^\rho 
\Gamma^i (\Gamma^j[\boldX^j, \boldPsi])D^\sigma \boldX^i
\boldF_{\rho\sigma} = 0
\eea
for the values of the coefficients given in  \eref{fermioncoeff},  $\hat {\bf d} =\sfrac{i}{4}$, $\hat {\bf e} =-\sfrac{i}{4}$,
$\hat {\bf g} = -\sfrac{i}{2} $, $\hat{\bf h} =-i$, $\hat
{\bf j} =i $ and $\hat {\bf m} =-\sfrac{2i}{3}$.

\section{Uniqueness of the four-derivative fermion ansatz }\label{Appunique}

When dealing with the fermionic part of the action one might worry
about the uniqueness claim of our proposal, since it looks as if there
are many additional terms that could lead to the operators present in
the $\alpha'^2$-corrected D2-brane action upon Higgsing. In order to
address that, we give below the most general set of expressions
obtained by `uplifting' the terms containing fermions in the D2 action
at order $\alpha'^2$. The `uplifting' procedure involves writing down
the most general 3-algebra expression that could reduce to a
particular D2 term by Higgsing.  The list excludes `on-shell' terms,
that is $\Gamma^\mu D_\mu \Psi$ and $\Gamma^{IJ}[X^I,X^J,\Psi]$, which
we will set to zero by using the lowest order 3-algebra equations of
motion. These terms would also have led to on-shell-type terms in the
D2 theory, which we know are absent, so we can safely set their
coefficients to zero.

In the following, the terms that appear in the main part of this paper
have been identified. The ones that did not have been enumerated and
we will show why they do not contribute to
\eref{fermansatz}. Ignoring signs and numerical factors we have:
\bea
  \bar{\boldPsi}\Gamma^{\mu}D^{\nu}\boldPsi\bar{\boldPsi}
\Gamma_{\nu}D_{\mu}\boldPsi\rightarrow &&
  \bar{\Psi}\Gamma^{\mu}D^{\nu}\Psi\bar{\Psi}
\Gamma_{\nu}D_{\mu}\Psi  \sim \textrm{term}\; \hat {\bf b}\cr 
  \bar{\boldPsi}\Gamma^{i}D^{\nu}\boldPsi\bar{\boldPsi}
\Gamma_{\nu}[\boldX^{i},\boldPsi]\rightarrow &&
  \bar{\Psi}\Gamma^{IJ}D^{\nu}\Psi\bar{\Psi}
\Gamma_{\nu}[X^{I},X^{J},\Psi]  \sim \textrm{term}\; \hat {\bf c}\cr 
  \bar{\boldPsi}\Gamma^{i}[\boldX^{j},\boldPsi]\bar{\boldPsi}
\Gamma^{j}[\boldX^{i},\boldPsi]\rightarrow &&
  \bar{\Psi}\Gamma^{IM}[X^J,X^N,\Psi]\bar{\Psi}
\Gamma^{JN}[X^I,X^M,\Psi]  \sim \textrm{term}\; \hat {\bf a}\cr 
  &&\bar{\Psi}\Gamma^{IN}[X^J,X^N,\Psi]\bar{\Psi}
\Gamma^{JM}[X^I,X^M,\Psi] \\
  &&\bar{\Psi}\Gamma^{IM}[X^J,X^N,\Psi]\bar{\Psi}
\Gamma^{JM}[X^I,X^N,\Psi]\\
  \bar{\boldPsi}\Gamma_{\mu}
D^{\nu}\boldPsi D^{\mu}\boldX^lD_{\nu}\boldX^l\rightarrow 
  &&\bar{\Psi}\Gamma_{\mu}D^{\nu}\Psi D^{\mu}
\boldX^LD_{\nu}\boldX^L  \sim \textrm{term}\; \hat {\bf e} \cr
  \bar{\boldPsi}\Gamma^{i}D^{\nu}\boldPsi 
D^{\rho}\boldX^i \boldF_{\rho\nu}\rightarrow 
  && \bar{\Psi}\Gamma^{IJ\rho\nu\lambda}D^{\nu}\Psi 
D^{\rho}X^I D_{\lambda}X^{J}\sim \textrm{term}\; \hat {\bf d}\cr 
  \bar{\boldPsi}\Gamma^{i}
D^{\nu}\boldPsi \boldX^{il}
D_{\nu}\boldX^{l}\rightarrow 
  && \bar{\Psi}\Gamma^{IM}D^{\nu}\Psi X^{ILM}
D_{\nu}X^{L} \sim \textrm{term}\; \hat {\bf g}\cr
  \bar{\boldPsi}\Gamma_{\mu}[\boldX^{j},\boldPsi]
\boldF^{\mu\rho}D_{\rho}\boldX^{j}\rightarrow
  &&\bar{\Psi}\Gamma^{\rho\lambda}[X^J,X^I,\Psi]
D_{\lambda}X^{I}D_{\rho}X^J \sim \textrm{term}\; \hat {\bf i}\cr 
  &&\bar{\Psi}\Gamma^{\rho\lambda KL}[X^J,X^K,\Psi]
D_{\lambda}X^{L}D_{\rho}X^J \sim \textrm{term}\; \hat {\bf j}   \cr 
  \bar{\boldPsi}\Gamma^i [\boldX^j,\boldPsi]
D^{\rho}\boldX^iD_{\rho}\boldX^j\rightarrow 
  &&\bar{\Psi}\Gamma^{IM} [X^J,X^M,\Psi]
D^{\rho}X^ID_{\rho}X^J  \sim \textrm{term}\; \hat {\bf h} \cr 
  \bar{\boldPsi}\Gamma_{\mu}[\boldX^j,\boldPsi]
D^{\mu}\boldX^l\boldX^{lj}\rightarrow
  &&\bar{\Psi}\Gamma_{\mu}[X^J,X^M,\Psi]
D^{\mu}X^LX^{LJM} \sim \textrm{term}\; \hat {\bf l}\cr
  &&\bar{\Psi}\Gamma_{\mu}\Gamma^{MN}[X^J,X^M,\Psi]D^{\mu}X^LX^{LJN}  \\
  \bar{\boldPsi}\Gamma_{\mu\nu\rho}[\boldX^k,\boldPsi]
\boldF^{\mu\nu}D^{\rho}\boldX^k\rightarrow
  &&\bar{\Psi}\Gamma^{MN}[X^K,X^M,\Psi]D_{\rho}X^N
D^{\rho}X^K  \sim \textrm{term}\; \hat {\bf h} \cr 
  \bar{\boldPsi}\Gamma_{\mu\nu l}D_{\sigma}\boldPsi 
\boldF^{\mu\nu}D^{\sigma}X^l \rightarrow
  && \bar{\Psi}\Gamma^{\mu LM}D_{\sigma}\Psi D_{\mu}X^M 
D^{\sigma}X^L \sim \textrm{term}\; \hat {\bf d}\cr 
  \bar{\boldPsi}\Gamma_{\rho\mu j}D_{\sigma}\boldPsi 
D^{\mu}\boldX^{j}\boldF^{\rho\sigma}\rightarrow
  && \bar{\Psi}\Gamma^{\mu IJ}D^{\nu}\Psi D_{\nu}X^{I}
D_{\mu}X^{J}   \sim \textrm{term}\; \hat {\bf d}  \cr 
  \bar{\boldPsi}\Gamma_{\mu j \rho} [\boldX^k,\boldPsi]
D^{\mu}\boldX^jD^{\rho}\boldX^{k}\rightarrow 
  &&\bar{\Psi}\Gamma_{\mu\rho JM} [X^K,X^M,\Psi]
D^{\mu}X^JD^{\rho}X^{K}\sim \textrm{term}\; \hat {\bf j} \cr 
  \bar{\boldPsi}\Gamma_{\mu jl}D_{\sigma}\boldPsi 
D^{\mu}\boldX^{j}D^{\sigma}\boldX^l \rightarrow 
  &&\bar{\Psi}\Gamma_{\mu JL}D_{\sigma}\Psi 
D^{\mu}X^{J}D^{\sigma}X^L  \sim \textrm{term}\; \hat {\bf d}\cr 
  \bar{\boldPsi}\Gamma_{\mu jl}[\boldX^k,\boldPsi]
D^{\mu}\boldX^j \boldX^{lk}\rightarrow
  && \bar{\Psi}\Gamma_{\mu JL}[X^K,X^{M},\Psi]
D^{\mu}X^J X^{LKM}  \sim \textrm{term}\; \hat {\bf k} \cr 
  && \bar{\Psi}\Gamma_{\mu}\Gamma^{JL}\Gamma^{MN}[X^K,X^{M},\Psi]
D^{\mu}X^J X^{LKN} \\
  \bar{\boldPsi}\Gamma_{ij\rho}
D_{\sigma}\boldPsi \boldX^{ij}\boldF^{\rho\sigma} \rightarrow
  && \bar{\Psi}\Gamma^{\rho\sigma}\Gamma^{M}\Gamma^{IJK}
D_{\sigma}\Psi X^{IJK}D_{\rho}X^M   \\
  && \bar{\Psi}\Gamma^{\rho\sigma}\Gamma^{IJ}
D_{\sigma}\Psi X^{IJM}D_{\rho}X^M  \\
  \bar{\boldPsi}\Gamma^{ij}\Gamma_{\rho}[\boldX^k,\boldPsi]\boldX^{ij}
D^{\rho}\boldX^{k}\rightarrow
  && \bar{\Psi}\Gamma^{IJ}\Gamma_{\rho}[X^K,X^M,\Psi]X^{IJM}
D^{\rho}X^{K}  \sim \textrm{term}\; \hat {\bf h} \cr
  &&\bar{\Psi}\Gamma^{IJN}\Gamma_{\rho}
\Gamma^{M}[X^K,X^M,\Psi]X^{IJN}D^{\rho}X^{K}\sim \textrm{term}\; \hat  {\bf m}   \cr
  \bar{\boldPsi}\Gamma_{ijl}D_{\sigma}\boldPsi \boldX^{ij}
D^{\sigma}\boldX^{l}\rightarrow
  && \bar{\Psi}\Gamma^{IJK}\Gamma^{M}D_{\sigma}\Psi X^{IJK}
D^{\sigma}X^{M}  \sim \textrm{term}\; \hat {\bf f}\cr 
  \bar{\boldPsi}\Gamma^{i}[\boldX^j,\boldPsi]\boldX^{il}\boldX^{lj}\rightarrow 
  && \bar{\Psi}\Gamma^{IM}[X^J,X^K,\Psi]X^{ILM}X^{LJK}  \sim \textrm{term}\; \hat {\bf p} \cr
  &&\bar{\Psi}\Gamma^{IM}[X^J,X^M,\Psi]X^{ILK}X^{LJK} \\
  &&\bar{\Psi}\Gamma^{IM}[X^J,X^N,\Psi]X^{ILN}X^{LJM} \\
  &&\bar{\Psi}\Gamma^{IK}\Gamma^{MN}[X^J,X^N,\Psi]X^{ILK}X^{LJM} \\
  \bar{\boldPsi}\Gamma_{\mu\nu l}[\boldX^k,\boldPsi]\boldF^{\mu\nu}\boldX^{lk}\rightarrow 
  &&\bar{\Psi}\Gamma^{\mu}\Gamma^{LN}[X^{K},X^M,\Psi]
D_{\mu}X^NX^{LKM} \sim \textrm{term}\; \hat {\bf k} \cr
  &&\bar{\Psi}\Gamma^{\mu}\Gamma^{LN}[X^{K},X^M,\Psi]
D_{\mu}X^MX^{LKN}  \sim \textrm{term}\; \hat {\bf n} \cr 
  &&\bar{\Psi}\Gamma^{\mu}\Gamma^{LM}[X^{K},X^M,\Psi]
D_{\mu}X^NX^{LKN}  \\
  &&\bar{\Psi}\Gamma^{\mu}\Gamma^{LN}\Gamma^{M}
\Gamma^{P}[X^{K},X^M,\Psi]D_{\mu}   X^PX^{LKN}  \\
  \bar{\boldPsi}\Gamma_{ijl}[\boldX^k,\boldPsi]\boldX^{ij}\boldX^{lk}\rightarrow
  &&\bar{\Psi}\Gamma^{IJLN}[X^K,X^M,\Psi]X^{IJN}X^{LKM} \sim\textrm{term}\;\hat {\bf o}\cr
  &&\bar{\Psi}\Gamma^{IJL}\Gamma^{N}[X^K,X^N,\Psi]X^{IJM}X^{LKM} \\
  &&\bar{\Psi}\Gamma^{IJL}\Gamma^{N}[X^K,X^M,\Psi]X^{IJM}X^{LKN} \\
  &&\bar{\Psi}\Gamma^{IJL}\Gamma^{N}\Gamma^{P}\Gamma^M[X^K,X^M,\Psi]X^{IJN}X^{LKP}
\eea
The enumerated terms do not contribute as they are either related to
terms already present in the ansatz (up to `on-shell' terms) or Higgs
to terms not present in the D2 theory and should therefore have a zero
coefficient. We have used the following $\epsilon$-tensor identity
 in showing the equivalence of several terms by re-shuffling
SO(8) indices amongst products of 3-brackets:
\begin{equation}
\epsilon^{a[bcd}\epsilon^{e]fgh} = 0\;,
\end{equation}
where the above indices are gauge indices and one should also remember
that there is a STr in front of each expression. This leads to
the fermionic analogues of \eref{euclidid}, the origin of which also
lies in the above identity and the implementation of the $\STr$
prescription. In more detail we have:
\begin{itemize}
\item (C.1) gives an on-shell term upon setting $I=J=8$
\item (C.2) gives a term that doesn't exist in D2 for $N=8\neq M$
\item (C.3) is equivalent to $\hat{\bf n}$
\item (C.4) Higgses to a term not present in D2 for $K=8$
\item (C.5) Higgses to a term not present in D2 for $M\neq 8$
\item (C.6) by expanding $\Gamma^{IJ} = \Gamma^I \Gamma^J - \delta^{IJ}$ reduces to $\hat {\bf g}$ and an on-shell term
\item (C.7) is equivalent to $\hat{\bf p} $ up to an on-shell term
\item (C.8) is equivalent to $\hat{\bf p} $ up to an on-shell term
\item (C.9) is equivalent to $\hat{\bf o} $ up to an on-shell term
\item (C.10) is the same as (C.5)
\item (C.11) Higgses to a term not present in D2 for $K=8$
\item (C.12) is equivalent to $\hat {\bf o} $ up to an on-shell term
\item (C.13) is equivalent to $\hat {\bf o} $ up to an on-shell term
\item (C.14) Higgses to a term not present in D2 for $K=8$

\end{itemize}
Therefore, the only independent terms are the ones with coefficients $\hat{\bf a} ,..., \hat {\bf p} $ that we have already included in \eref{fermansatz}.

\end{appendix}

\bibliographystyle{JHEP}
\bibliography{F4}

\end{document}